\documentclass[preprint,11pt,authoryear,times]{elsarticle}

\usepackage[utf8]{inputenc}
\usepackage[english]{babel}
\usepackage{pdflscape}
\usepackage{graphicx}
\usepackage{subcaption}
\usepackage{setspace}
\usepackage[ruled,vlined]{algorithm2e}
\usepackage{mwe}
\usepackage{a4wide}
\usepackage{float}
\usepackage{booktabs}
\usepackage{amsfonts}
\usepackage{amsmath}
\usepackage{amsthm}
\usepackage{xcolor}
\usepackage{hyperref}            
\usepackage{lineno}
\usepackage[authoryear]{natbib}
\usepackage{multirow} % 
\usepackage{hhline}   % 
\usepackage{comment}
\usepackage{ulem}
\usepackage{subcaption}
\usepackage[font=small,skip=0pt]{caption}
\modulolinenumbers[5]  
\hypersetup{%
    bookmarksnumbered,
    pdfstartview={FitH},
    linkcolor=blue,
    citecolor=blue,
    urlcolor=blue,
    colorlinks=true,
    plainpages=false%,
}
\newtheorem{defn}{Definition}

\makeatletter
\def\@author#1{\g@addto@macro\elsauthors{\normalsize%
    \def\baselinestretch{1}%
    \upshape\authorsep#1\unskip\textsuperscript{%
      \ifx\@fnmark\@empty\else\unskip\sep\@fnmark\let\sep=,\fi
      \ifx\@corref\@empty\else\unskip\sep\@corref\let\sep=,\fi
      }%
    \def\authorsep{\unskip,\space}%
    \global\let\@fnmark\@empty
    \global\let\@corref\@empty  %% Added
    \global\let\sep\@empty}%
    \@eadauthor={#1}
}
\makeatother
\begin{document}
\begin{frontmatter}
\journal{Computational Statistics and Data Analysis}
\title{Simultaneous predictive bands for functional time series using minimum entropy sets}

\author{Nicolás Hernández\fnref{fn1}\corref{cor1}}
\author{Jairo Cugliari\fnref{fn2}}
\author{Julien Jacques\fnref{fn2}}
\address[fn1]{Department of Statistical Science, University College London, UK}
\address[fn2]{Laboratoire ERIC ER 3083, Université Lumière - Lyon 2, France}
\cortext[cor1]{Email: n.hernandez@ucl.ac.uk}

\begin{abstract}

Functional Time Series (FTS) are sequences of dependent random elements taking values on some functional space. Most of the research on this domain focuses on producing a predictor able to forecast the next function, having observed a part of the sequence. For this, the Autoregressive Hilbertian process is a suitable framework. Here, we address the problem of constructing simultaneous predictive confidence bands for a stationary FTS. The method is based on an entropy measure for stochastic processes. To construct predictive bands, we use a Reproducing Kernel Hilbert Spaces (RKHS) to represent the functions and a functional bootstrap procedure that allows us to estimate the prediction law and a Reproducing Kernel Hilbert Spaces (RKHS) to  represent the functions, considering then the basis associated to the reproducing kernel. We then classify the points on the projected space according to those that belong to the minimum entropy set (MES) and those that do not. We map the minimum entropy set back to the functional space and construct a band using the regularity property of the RKHS. The proposed methodology is illustrated through artificial and real data sets.
\end{abstract}

% \begin{keyword}
% Functional time series \sep Autoregressive Hilbertian process \sep RKHS \sep Bootstrap \sep Entropy \sep Predictive bands. 
% \end{keyword}

\date{\today}

\end{frontmatter}

%%%%%%%%%%%%%%%%%%%%%%%%%%%%%%%%%%%%%%%%%%%%%%%%%%%%%%%%%%%%%%%%%%%%%%%%%%%%%%%%%%%%%%%%
%%                                                                                    %%
%%                                  S  E  C  T  I  O  N                               %%
%%                                                                                    %%
%%%%%%%%%%%%%%%%%%%%%%%%%%%%%%%%%%%%%%%%%%%%%%%%%%%%%%%%%%%%%%%%%%%%%%%%%%%%%%%%%%%%%%%%

\section{Introduction}\label{ss:intro}

While Functional Data Analysis  \cite{ramsay2006functional} deals with independent and identical distributed realizations of functional data, the term Functional Time Series (FTS) refers to dependent series of random elements lying in some functional space. Alongside the academic research interest awoken by such an elegant framework, FTS have important practical usefulness. In some cases, they allow one to consider an alternative modelling scheme to classical time series, thanks to their ability to intrinsically deal with non-stationary patterns such as seasonality. Moreover, they can be of great help if one has a heterogeneous sampling scheme where records are observed at unequally spaced time points.

A simple yet powerful device to construct a FTS is proposed by \cite{bosq2000linear}. The idea is to take slices of an underlying continuous stochastic process, say $\xi = \lbrace \xi(t); t\in\mathbb{R}\rbrace$, using a window of fixed size $\delta > 0$. We then consider the FTS $Z$ such that $Z_k(t) = \xi((k - 1)\delta + t)$ with $k\in\mathbb{Z}$ and $t\in(0, \delta)$. In many real data applications, this scheme is reasonable: an underlying phenomenon of continuous nature exists even if data are recorded at some sampling rate. Examples of this can be found, for example, in electrical device monitors \citep{bonnevay2019predictive}, energy demand \citep{ANTONIADIS2016939}, air pollution \citep{paparoditis2020bootstrap} and geology \citep{hormann2010weakly}. Note that the device is particularly fruitful if the quantity $\delta$ is tailored to some seasonal component of the series $\xi$. In such cases, the series $Z$ naturally incorporates the non-stationary pattern induced by the seasonality. Other kinds of FTS involve more general situations, as for instance sequences of functions that do not verify continuity constraints between consecutive functions (see \citet{aue2015FTS}).

One important task related to FTS is forecasting. That is, after having observed $n$ realizations of the process $Z$, say $Z_1, \ldots, Z_n$, the aim is to predict the next whole function $Z_{n+1}(t), t\in(0, \delta)$. Autoregressive Hilbertian processes (ARH) are the natural extension of univariate autoregressive processes to the functional data framework where Hilbert spaces are considered. The term Functional Autoregressive is sometimes also used to name the same process, while this can produce confusion with univariate autoregressive non-linear processes. In short, they consist of linking two consecutive functions of a stationary FTS through a linear operator plus a functional noise. Since the functions are usually supposed to take values on some Hilbert space, the processes are also named Autoregressive Hilbertian (ARH). An excellent review of these processes can be found in \citet{alvarez2017review}. Several works address the problem of predicting the function $Z_{n+1}(t)$ using variants of ARH \citep{aue2015FTS, nagbe2018short, wang2020functional}; in a more general framework such as a non-linear version of FTS \citep{antoniadis2006functional}, or using Factor Models \citep{tavakoli2023factor}.

While point prediction gives helpful information about the future evolution of FTS, uncertainty quantification is often needed to produce valid inferences. Estimating the predictive law has received much less attention in the FTS literature. In general, bootstrap methods are used to produce pseudo-predictions. Some form of central region is then taken to be the predictive band. For instance, \citet{hyndman2009forecasting} describe the FTS as a set of univariate time series associated to projected components. Then, using classical time series tools, they produce individual predictions and pseudo-predictions based on a residual bootstrap. The univariate pseudo-predictions are mapped back to the functional space to obtain functional pseudo-predictions. With this, the band is constructed on a point-by-point basis taking the quantiles at some coverage level. \cite{antoniadis2012} use a non-linear version of FTS where the predictor is a weighted mean of functions with weights increasing with the similarity between the last observed function and every function in the dataset. The set of weights is the key element of the bootstrap procedure since the more a function is similar with the current one, the more likely the chances of its next function are to be sampled. The band is constructed using heuristics coming from econometric literature. In \citet{paparoditis2020bootstrap} the authors propose a model-free bootstrap procedure based on the prediction generated by a vector autoregressive model (VAR) applied to the coefficients of the Karhunen-Lo\`eve expansion of the FTS. Then simultaneous and pointwise intervals for the prediction are constructed using the bootstrap studentized prediction error.

In this paper we address the problem of constructing a confidence band for the prediction of the FTS. The aim is to construct a region $\mathcal{B}^{1-\alpha}_{n+1}$ which covers the function $Z_{n+1}(t)$, that is $P (Z_{n+1}(t) \in \mathcal{B}^{1-\alpha}_{n+1}) \geq 1 - \alpha$ for a fixed risk level $\alpha \in (0,1)$. The probability measure is found using the information available at the moment $n$. Note that this prediction band requires simultaneous coverage of the whole curve $Z_{n+1}(t)$, which, in general, is a difficult requirement (see \citet{degras2017simultaneous} for a discussion in the functional data context). Since it is usual to represent functions by the projection coefficients over a basis, some works have explored the option of constructing a confidence region on the coefficients space \citep{antoniadis2006functional, antoniadis2012}. However, obtaining simultaneous confidence bands sometimes requires additional regularity assumptions on the functional spaces \citet{Genovese2005}.

Probability densities play an important role in the construction of predictive intervals. However, for functional data the concept of density cannot be well defined in the functional space. In this sense, as the usual workflow is designed to deal with the projected coefficients of the functions in some basis \citet{delaigle2010defining} propose that we consider the probability density in such multivariate space. The construction of probability regions in the projected space can be sorted out by means of a metric e.g., a distance that induces an order in that space. This metric can be a depth function, (\citet{fraiman, lopez2009concept, cuevas2007robust}, among others); a distance (\citet{galeano2015mahalanobis, martos2014generalizing, cardot2013efficient} and the references therein); or an information theory tool such as entropy (\citet{martos2018entropy}). These three concepts are linked in the sense that they induce an order in a functional data set and allow the construction of $(1-\alpha)$-central regions in the original space where the functions inhabit. Ideally, the main condition that these $(1-\alpha)$-central regions should satisfy is that they concentrate a high amount (at least $1-\alpha\text{, with } \alpha\in [0,1]$) of probability inside. In multivariate spaces, these $(1-\alpha)$-central regions are known in the literature as high density regions (HDR), see \citet{hyndman1996computing}.

In this paper we propose to use Minimum Entropy Sets (MES) as a way to estimate the HDR of the predictive law of the projection coefficients to then construct prediction bands. Our procedure can be used on top of any FTS forecasting approach, providing a predictor and a bootstrap method. Additional regularity of the functional space may be required to map the MES back on to the functional space. Our choice is to work on the special case of Hilbert spaces given by reproducing kernels (RKHS) that ensures the continuity of the pointwise evaluation functional.

\begin{figure}[!ht]
\centering
\caption{Band construction procedure.}
\includegraphics[width=\textwidth]{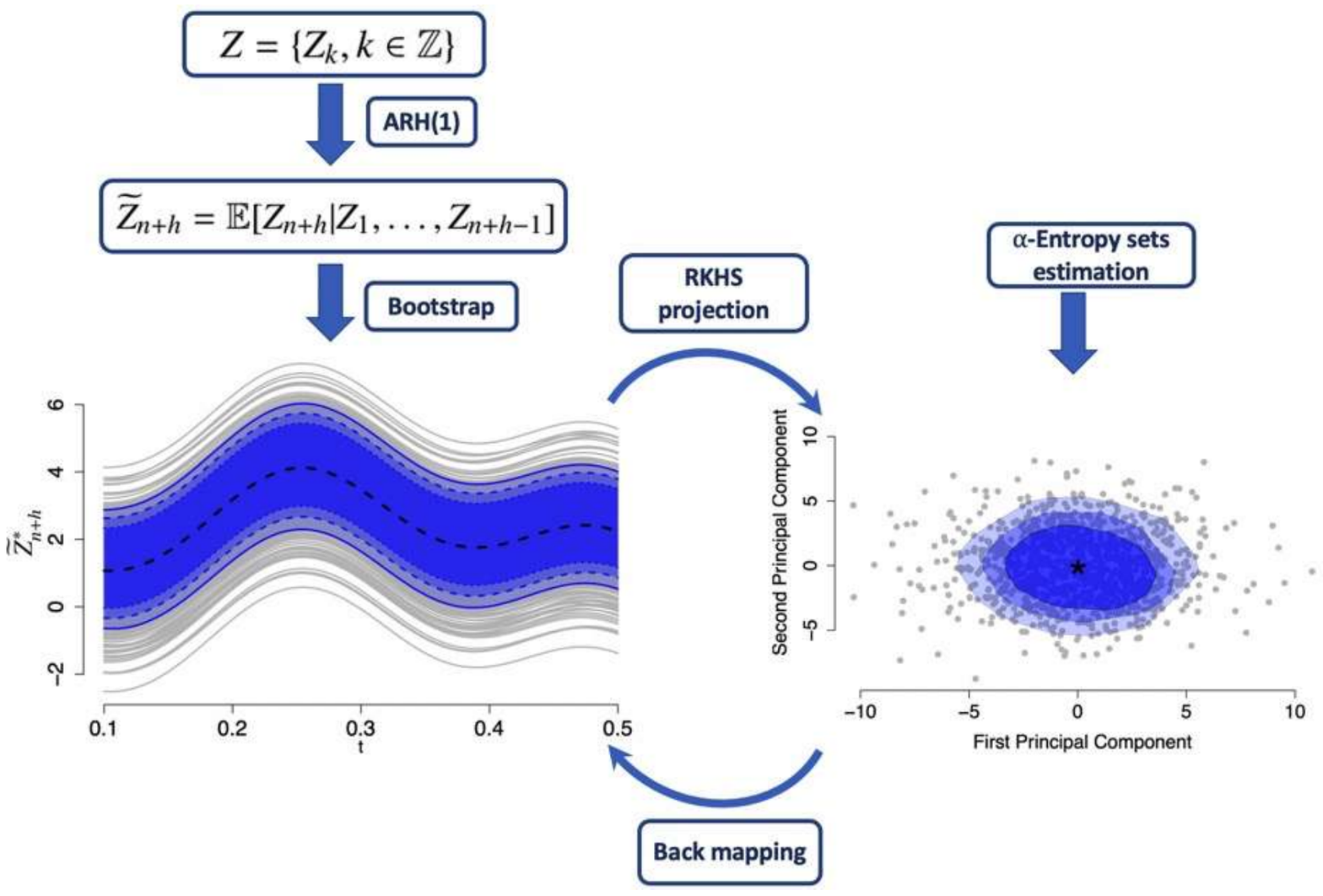}
\label{Ex1}
\end{figure}

We illustrate our procedure in Figure \ref{Ex1}. Departing from a functional time series $Z$, we obtain the h-step ahead prediction using, for example, an ARH(1) functional model. Using the residual based functional bootstrap procedure presented in Section \ref{bootstrap} we obtain the sample of predictive pseudo-replicates. These are represented as grey curves in the background of the leftmost graphic. The functions are projected into an appropriate space $\mathcal{H}_k\subset\mathcal{H}$ (defined in Section \ref{ss:rkhs}) and now work on the coefficients of that projection (rightmost graphic). Then, we compute the MES of this new space, for different values of confidence $1-\alpha$ (in the example we use $\alpha=\{0.05,0.1,0.2\}$). The final step involves the back-mapping of these sets to the functional space. For this, we identify each point covered by the MES on the right with its associated curve on the left. Finally, the predictive confidence band is the convex-hull of the curves associated to the MES.

The paper is organized as follows: in Section \ref{sec:ARH} we present the RKHS framework needed for representing the ARH processes in such spaces, as well as considerations about the simulation, estimation and prediction procedures. We focus on predictive inference in Section \ref{sec:predictive}, defining Minimum Entropy Sets, a bootstrap procedure compatible with FTS and the predictive confidence bands methodology. Section \ref{sec:experiments} describes the experimental design using Monte Carlo simulation  to show the good covering results of our proposal and a real data application. We conclude the paper with Section \ref{sec:conclusion}.

%%%%%%%%%%%%%%%%%%%%%%%%%%%%%%%%%%%%%%%%%%%%%%%%%%%%%%%%%%%%%%%%%%%%%%%%%%%%%%%%%%%%%%%%
%%                                                                                    %%
%%                                  S  E  C  T  I  O  N                               %%
%%                                                                                    %%
%%%%%%%%%%%%%%%%%%%%%%%%%%%%%%%%%%%%%%%%%%%%%%%%%%%%%%%%%%%%%%%%%%%%%%%%%%%%%%%%%%%%%%%%

\section{The autoregressive Hilbertian model: ARH}\label{sec:ARH}

In this section we introduce notation and recall well-known facts about representing a function with Reproducing Kernel Hilbert Spaces (RKHS) and autoregressive Hilbert (ARH) processes. All the random elements below are defined in the same probability space $(\Omega,\mathcal{F},P)$. 

% --------------------------------------------------------------------------------------
\subsection{Representing functional data using RKHS}\label{ss:rkhs} % 

In general, we cannot observe the full trajectories of functional variable realisations. Since it is also the case for FTS, our input data is the set of discrete and noisy trajectories that constitutes realisations of the functional variable, also known in the literature as raw functional data \citep{hsing2015theoretical}. In what follows, we set the functional space $\mathcal{H} \subset C[T]$ the space of (classes of) real continuous functions over the compact T -. In our case we set $T = [0, \delta]$, with $\delta = 1$ without loss of generality. In that sense, raw functional data consist of the collection of the functions recorded over a grid of $m$ discretized points, say $t_1,\dots,t_m$ (usually equally spaced). The analysis departs from a collection of discrete functions, hereafter curves: $\{z_j(t_i)\}_{i=1}^m\text{ for } j=1,\dots,n$.\\

As is usual in the functional data literature, in order to estimate the underlying functional object that generated each realisation $z_j$ we need to choose a system of orthonormal basis functions. Following the approach in \cite{munoz2010representing, berlinet2011reproducing}, we propose to chose $\mathcal{H} = \mathcal{H}_K $ as a RKHS, such that the family of basis functions $\Phi=\{\phi_1,\dots,\phi_d\}$ generates the functional subspace $\mathcal{H}_K$ through its linear span. These basis functions are linked to the positive-semidefinite kernel function $K = K(s, t)$ associated to $\mathcal{H}_K$. Any realisation $z(t)$ can be approximated by the following functional estimator
\begin{equation}\label{Eq.5}
\tilde{z}(t):=  \operatorname{arg} \min_{f\in \mathcal{H}} 
  \sum^{m}_{i=1}L(z(t_i),f(t_i))
  + \gamma \|f\|_{\mathcal{H}}^2, 
\end{equation}
where $\gamma > 0$ is a regularization parameter, $\|f\|_{\mathcal{H}}$ is the norm of the function $f$ in $\mathcal{H}$ and $L(w,z) = (w-z)^2$ is a loss function. By the Representer Theorem \cite[Theorem~5.2, p.~91]{smale} the solution of the problem stated in Eq.~\eqref{Eq.5} exists, is unique, and admits the following representation
\begin{equation}\label{Eq.6}
\tilde{z}(t)=\sum_{i=1}^m a_{i} K(t_i,t) = \textbf{a}^T \mathbf{K}_t,
\end{equation}

\noindent
where $\mathbf{K}_t = \{K(t_1,t),\dots,K(t_m,t)\}$ is the vector of kernel evaluations and the linear combination coefficients $\textbf{a} = (a_{1},\dots,a_{m}) \in \mathbb{R}^m$ are obtained as the solution to the linear system $(\gamma \mathbf{I}_m + \textbf{K})\textbf{a} = \mathbf{z}$, for $\mathbf{z} = (z(t_1), \dots, z(t_m))^T$, $\mathbf{I}_m$ an $m \times m$ identity matrix, and $\textbf{K}$ the Gram matrix with the kernel function evaluations over the grid $t_1, \dots, t_m$. Now, we use the Mercer decomposition theorem \citep{j1909xvi} to obtain the basis. Indeed, since K is a positive-definite and symmetric kernel function with associated integral operator $I_K(z) = \int_T K(\cdot,t)z(t)dt$, it admits a spectral decomposition into a sequence $(\lambda_i,\phi_i)_{i \geq 1}$ of eigenvalue-eigenfunction pairs. 
Note that in this scheme, one chooses the kernel K, which determines the basis, which is the converse of usual FDA practice. Then, each estimated functional datum $\tilde{z}(t)$ in the sample can be expressed as follows
\begin{equation}\label{Eq.7}
\tilde{z}(t)=\sum_{i=1}^{m} c_{i} \phi_i(t),
\end{equation}
where $c_{i}$ are the projection coefficients of $\tilde{z}(t)$ onto the space generated by the eigenfunctions $\phi_i$. Nevertheless, the expression in Eq.~\eqref{Eq.7} is an unhelpful representation when the sequence of eigenpairs $(\lambda_i, \phi_i)_{i \geq 1}$ is unknown. With the sample data at hand, $c_{i}$ can be estimated by:
\begin{equation}\label{lambdas}
    \hat{c}_{i} = \displaystyle\frac{l_i}{\sqrt{m}}(\textbf{a}^T\mathbf{v}_i),
\end{equation}

\noindent
where $(l_i, \mathbf{v}_i)$ are the $i^{th}$ eigenvalue (in decreasing order) and eigenvector of $\textbf{K}$. It is possible to approximate the development in Eq.~\eqref{Eq.7} by truncating the sum by the first few terms. Therefore, we represent each function $z(t)$ with a finite representation in $\mathbb{R}^d$ given by $\hat{\pmb{c}} = \{\hat{c}_{1},\dots,\hat{c}_{d}\}$, where $d \leq \text{rank}(\textbf{K}$). Then, the function bases are truncated and the projection coefficients approximated via equation \eqref{lambdas}. The dimension $d$ is simply a truncation order, not something intrinsic to the kernel $K$ or space $\mathcal{H}$.

% --------------------------------------------------------------------------------------
\subsection{Autoregressive Hilbert process}
Following the previous notation, the sequence $Z = \{Z_k, k \in \mathbb{Z}\}$ of random functions on $\mathcal{H}$ follows an ARH(1) if,
\begin{equation}\label{eq:arh}
  Z_k   = \mu + \Psi (Z_{k-1} - \mu ) + \epsilon_k,
\end{equation}
\noindent
where $\mu$ is the mean function, $\Psi$ is a linear bounded (continuous) operator on $\mathcal{H}\mapsto \mathcal{H}$ and $\epsilon = \lbrace \epsilon_k, k \in \mathbb{Z} \rbrace$ a strong white noise in $\mathcal{H}$ such that $\mathbb{E}\|\epsilon_k\|< \infty$ and $E [Z_0] = \mu$, $E \|Z_0\|^2 < \infty$. Under mild conditions, the expression defined in Eq.~\eqref{eq:arh} defines a strictly stationary stochastic process in $\mathcal{H}$. Given $Z_1, \ldots, Z_n$, the best predictor of the function $Z_{n+1}$ in the mean-square-error sense is given by 
\begin{equation}\label{eq:arhpredictor}
\widetilde{Z}_{n + 1} = \mathbb{E} [Z_{n+1}|Z_1, \ldots, Z_n],
\end{equation}

\noindent
which results for the ARH process in $\widetilde{Z}_{n + 1} = \mu + \Psi(Z_n - \mu)$. This predictor is not statistical in nature because it depends on parameters of the unknown probability law. Therefore, one needs to estimate $\mu$ and $\Psi$ to get a statistical predictor. A simple way is to use the empirical counterpart of the population mean to estimate the mean function. The estimation of $\Psi$ is more cumbersome because it involves unbounded operators. For this, the cross-covariance operator of lag $r$ is defined as
\begin{equation}\label{eq:gammar}
\Gamma_r = \mathbb{E} [(Z_0 - \mu) \otimes (Z_r - \mu)],
\end{equation}
which under mild conditions is a trace-class operator. To estimate the operator $\Psi$ we use the covariance operator ($\Gamma_0$) and the cross-covariance operator ($\Gamma_1$), related by a Yule-Walker like expression (see \cite{bosq2000linear}), 
\begin{equation}\label{eq:yule-walker}
  \Gamma_0 \Psi = \Gamma_1.
\end{equation}

If $\mathcal{H}$ is of finite dimension, then one can just plug in the empirical counterparts of $\Gamma_0$ and $\Gamma_1$ and solve to obtain $\Psi$. However, in the general case the unboundedness of the inverse of $\Gamma_0$ implies that the expression in Eq.~\ref{eq:yule-walker} is an ill-posed problem, see \citet{mas2011linear} for a detailed discussion. Fortunately, the problem has more impact for the development of theoretical results. The empirical equivalents of the operators are matrices and one relies on regularized versions of the inverse of the covariance matrix to obtain the estimation of $\Psi$. A second important property of ARH processes is worth mentioning since it is used for simulation. Indeed, if $Z$ follows an ARH(1), then 
\begin{equation}\label{eq:arh-simu}
  \Gamma_\epsilon =  \Gamma_0 - \Psi \Gamma_0 \Psi^T,    
\end{equation}
where $\Gamma_\epsilon$ is the covariance operator of $\epsilon$ \citep[Chap 3.]{bosq2000linear}. This means that only two of the three operators can be chosen freely for the simulation of ARH trajectories. Moreover, since $\Gamma_\epsilon$ needs to be positive-definite, not every couple of $\Psi$ and $\Gamma_0$ will produce a compatible innovation process $\epsilon$.

\paragraph{Estimation and prediction} In practice, only finite measurements on each function are used to estimate the parameter. Given the finite, truncated representation of each curve in Eq.~\ref{Eq.7} and the estimated coefficients, $\hat{c}_i$, the mean function can be estimated by
\begin{equation}\label{eq:estim-mean}
\hat{\mu} (t) =  \sum_{i=1}^d \bar{\hat{c}}_{i} \phi_i (t),
              \qquad \text{with} \;\;  \bar{\hat{c}}_{i} = \frac{1}{n} \sum_{k=1}^n \hat{c}_{k,i}.
\end{equation}
As it is usual in FDA the operation concerns the projection coefficients, which are then expanded on the basis constructed using a kernel function. Similarly, the corresponding covariance $\hat\Gamma_0$ and cross-covariance $\hat\Gamma_1$ estimators of the operators $\Gamma_0$ and $\Gamma_1$ are 
\begin{equation}\label{eq:estim-cov}
\widehat{\Gamma}_r = \frac{1}{n-r} \sum_{k=1}^{n-r} \sum_{i'=1}^d 
    \sum_{i=1}^d \hat{d}_{k,i} \hat{d}_{k,i'+r} \phi_i \otimes \phi_{i'},
    \qquad r = 0, 1, 
\end{equation}
where $\hat{d}_{k,i} = \hat{c}_{k, i} -\bar{\hat{c}}_{i}$. Note that the coefficients can be arranged into two squared $d\times d$ matrices, say $\mathbf{\hat{C}_0}$ and $\mathbf{\hat{C}_1}$,
\begin{equation}\label{eq:estim-matcov}
(\mathbf{\widehat{C}_r})_{i,i'} = \frac{1}{n-r} \sum_{k=1}^{n-r} \hat{d}_{k,i} \hat{d}_{k,i'+r} ,
    \qquad r = 0, 1. 
\end{equation}
At this point, one can estimate the projection coefficients of $\Psi$ on the double basis induced by the kernel. For this, we solve the equivalent of Eq.~\eqref{eq:yule-walker} to obtain 
$\mathbf{\widehat{P}} = \mathbf{\widehat{C}_0^{-1}} \mathbf{\widehat{C}_1} $. Note that the truncation introduced in Section \ref{ss:rkhs} sized down these matrices to $d<m$. Besides the trivial reduction on the computation times of these matrices, the truncation induces a regularization effect needed to better approximate the operator $\Psi$ (see \citet{bosq2000linear}). 
Finally, $\Psi$ is estimated by:
\begin{equation}\label{eq:estim-psi}
  \widehat{\Psi} = \sum_{i'=1}^d 
    \sum_{i=1}^d \hat{p}_{i, i'} \phi_i \otimes \phi_{i'},
\end{equation}
where $\hat{p}_{i, i'} = (\mathbf{\widehat{P}})_{i,i'}$. The one-step-ahead prediction $\widehat{Z}_{n+1|n}$ is obtained by applying $\widehat{\Psi}$ to the last observed function.

\paragraph{Simulation} We adopt a straightforward approach to simulate an ARH as in \cite{didericksen2012empirical}. Following equation \eqref{eq:arh}, the idea is to sequentially produce a FTS by applying the iteration after an arbitrary initialisation. By the ergodicity, one can expect that the obtained FTS follows the given ARH  after a reasonable burn-in period. 
This scheme is particularly fruitful to test different variants for $\mu$ and specifically for $\Psi$ (for instance symmetric and asymmetric operators). Moreover, one can choose different noise processes, which is important in the assessment of prediction bands. Importantly to notice, the explicit choices of the noise process (and its covariance operator $\Gamma_\epsilon$) and the autoregression operator $\Psi$ determine the variance operator $\Gamma_0$ of the FTS $X$.

\section{Making predictive inference with the ARH-RKHS model}
\label{sec:predictive}

Prediction is one of the main objectives in time series analysis. In terms of uncertainty, prediction intervals bring more information about the future values of a random variable than point
forecasts. In that sense, prediction intervals entail a set of values that the realisation of the future random variable could take, conditional on past information and given a certain probability. In the functional time series context, given the functional nature of the observations, the predictive confidence intervals take the form of predictive confidence bands (PCB), which implies the definition of a bounded region in the original space, $\mathcal{H}$, which the functions inhabit, such that if we randomly take one realisation of the sample of functional time series, it will be fully enclosed by the band with a given probability.

We define the lower and upper functional statistics, $L_{n+1|n}(t)$ and $U_{n+1|n}(t)$ both in $\mathcal{H}$ such that the region comprised by $\{[L_{n+1|n}(t),U_{n+1|n}(t)]:t\in T\}$ fully contains the conditional expectation  $\widetilde{Z}_{n+1|n}$ (see Eq.~\eqref{eq:arhpredictor}) with a probability of $1-\alpha$. 
\begin{defn}\label{banddef}
Let $L_{n+1|n}(t)$ and $U_{n+1|n}(t)$ be two elements of $\mathcal{H}$. We define a predictive confidence band as the set $\mathcal{B}^{1-\alpha}_{n+1|n} = \{(t, z) \in R^2 : L_{n+1|n}(t) \leq z \leq U_{n+1|n}(t), t\in T\}$, such that 
\begin{equation}\label{Eq.13}
P\big(\widetilde{Z}_{n+1|n} \in \mathcal{B}^{1-\alpha}_{n+1|n}, \forall t\in T\big) \geq 1- \alpha.
\end{equation}
\end{defn}

Following \citet{degras2017simultaneous}, there are at least two options to conduct predictive inference in the functional context: i) pointwisely or ii) simultaneously. Under the pointwise estimation of $\mathcal{B}^{1-\alpha}$ the condition stated in Eq.~\ref{Eq.13} is satisfied for each $t \in T$, independently. Let say the prediction is sampled at $m$ points in the domain $T$, then the pointwise confidence band, $\mathcal{\widehat{B}}^{1-\alpha,p}_{n+1|n}$ is constructed by the concatenation of $m$ prediction intervals $[L_{n+1|n}^p(t_i), U_{n+1|n}^p(t_i)]$ for each $t_i$, with $i=1,\dots,m$\footnote{Supraindex $p$ states for pointwise.}. The independent assumption implies that the joint probability is equal to the multiplication of the marginal probabilities for each $t_i$, then
\begin{equation*}
P\big(L_{n+1|n}^p(t_i) \leq \widetilde{Z}_{n+1|n} (t_i) \leq U_{n+1|n}^p(t_i), \forall i = 1,\dots,m) = (1- \alpha)^m \leq (1-\alpha), 
\end{equation*}

\noindent
which does not satisfy the condition stated in Eq.~\ref{Eq.13} (with the exception of trivial cases). Even though the pointwise predictive bands are a valid inferential method, in general their coverage is less than $1-\alpha$, which can mislead the conclusion in terms of the confidence of the prediction, see \cite{degras2017simultaneous} for a deeper discussion. To ensure that the condition stated in Eq.~\ref{Eq.13} is satisfied, which means that the coverage of the band is at least $1-\alpha$ for the entire domain, we need to tackle this problem under a simultaneous approach. Our choice is to approximate $\mathcal{B}^{1-\alpha}_{n+1|n}$ using $1-\alpha$ Minimum Entropy Sets (MES) on the $d$-dimensional representation space of functions as defined in Section \ref{ss:rkhs}. 

% ----------------------------------------------------------------------------
\subsection{$1-\alpha$ minimum entropy sets}

A tool that is of great help for the construction of PCB is the definition of a $1-\alpha-$Region, which is the minimum set that accounts for a given level of probability $\alpha \in [0,1]$. A common way to estimate this $1-\alpha-$Region is by means of a metric that orders the data sample with respect to a location statistic, typically the mean or the median. This metric can be a distance, a depth measure or a more complex function. Depth measures are widely used in functional data analysis, which were first studied by \cite{fraiman}. Other alternatives were introduced by \cite{lopez2009concept}, \cite{chakraborty2014data} and  \cite{cuevas2007robust}. The aforementioned depth measures present different drawbacks: i) \cite{fraiman} and \cite{lopez2009concept} use the original representation of the functional data - ($t_i, x(t_i)$) in the case of a set of univariate time series - ignoring the functional nature of the data. ii) The computation of these metrics requires evaluating many integrals, which is impractical when working with many curves. iii) Some of these measures, such as the random Tukey depth, that reduce computational cost by evaluating projections, do not provide a stable criterion to order the functions. iv) In the case of the modified band depth and the integrated depth, the coordinate-wise median is the curve with the highest depth (see \cite{chakraborty2014deepest} for a formal proof). This can lead to inaccurate results when there are non-linearities in the functional space where the curves are defined. 

In this paper, we consider an Information Theory criterion of minimum entropy. The entropy of a random variable gives information about the uncertainty associated to its realisations. In this sense, it constitutes an element that can be used to induce an order in the data and a tool to estimate the centre of a distribution and the outlyingness of its realisations. In the field of Information Theory the concept of entropy was introduced by \citet{shannon1948mathematical} and generalized by \citet{renyi1961measures}. Consider a random variable $C \in \mathbb{R}^d$ distributed according to a measure $F$ that admits a probability density function $f_C$. Then the \textit{R\'enyi} entropy aka Collision entropy, is computed as follows:

\begin{equation}\label{entropy}
H(C) = - \log \left( \int_{\mathbb{R}^d}f^{2}_C(\mathbf{c})d\mathbf{c}\right). 
\end{equation}

Let us consider the following example, when  $C \in \mathbb{R}^d$ is a multivariate normal random variable. For example, for a Gaussian distribution $C \sim \mathcal{N}_d(\mathbf{\mu}, \mathbf{\Sigma})$ the entropy can be computed as $H(C)=\displaystyle\frac{1}{2}\log(2\pi e)^d |\mathbf{\Sigma}|,$ where $d$ denotes the dimension of the space. We now consider the concept of $1-\alpha$-MES. This concept was introduced by \citet{hero2007geometric} and extended to the functional data framework by \citet{martos2018entropy}. Let $C \in \mathbb{R}^d$ a real-valued vector with continuous density function $f_C$ and a Borel set $A  \subset \mathbb{R}^d$ measurable with respect to the distribution function $F_C$; the entropy of the set $A$ with respect to the random vector $C$ is defined by $H(A,C) = -\log \left( \int_A f^{2}_C(\mathbf{c})d\mathbf{c}\right).$ Then, a set $\mathcal{R}_{1-\alpha}(C)$ is said to be the $1-\alpha$-MES for the random variable $C$, if satisfies the following optimization problem:

\begin{equation}\label{MES}
    \mathcal{R}_{1-\alpha}(C) :=  \operatorname{arg\, min}_{A \subset \mathbb{R}^d}H(A,C)  \text{ s.t. } P(A) \geq 1-\alpha,
\end{equation}

\noindent
for $0 < \alpha < 1$. To estimate the $\mathcal{R}_{1-\alpha}(C)$ one can assume a particular probability model for $F_C$. In order to obtain a more flexible and distribution-free estimator, we consider a non-parametric estimator using a local version of the entropy measure in Eq.~\eqref{entropy}. Given a positive proximity parameter $\delta$, consider the ball $\mathbb{B}_{(c,\delta)}$ centered at $\mathbf{c}$ such that $\delta = \int_{\mathbb{B}} f_C(\mathbf{c}) d\mathbf{c}$. The $\delta$-local entropy at the point $\mathbf{c}$ is defined as $H(\mathbb{B}_{(c,\delta)},C)$.

Given a set of data points $C_n=\{\mathbf{c}_1,\dots,\mathbf{c}_n\} \in \mathbb{R}^d$, the local entropy induces an order since it computes a univariate measure for each data point. The local entropy for the data at hand is obtained using the estimator $\widehat{H}(\mathbb{B}_{(\mathbf{c}_i,\delta)},C)=\exp(\bar{d}_k(\mathbf{c}_i,C_n))$, where $\bar{d}_k(\mathbf{c}_i,C_n)$ is the average distance from $\mathbf{c}_i$ to its \textit{k-th} nearest neighbours in $C_n$, according to \cite{martos2018entropy}. Note that the proximity parameter $\delta$ is defined implicitly by the number of nearest neighbours $k$. For the non-parametric estimation method to be consistent the speed of convergence needs to be lower than the linear case (see \cite{wand1994kernel} for further details). In particular, the authors in \cite{martos2018entropy} use $k= \sqrt{2n}$ as a rule of thumb, where they show that the results are robust with respect to the number of neighbours $k$ considered.

Now we explain how we estimate the $\mathcal{R}_{1-\alpha}(C)$ using the $\delta$-local entropy. A binary decision rule can be used to define whether a point belongs to the $\mathcal{R}_{1-\alpha}(C)$ or not by simply taking the points that are below the $1-\alpha$ quantile of the sets of estimated local entropies. This rule is theoretically justified in \cite{munozestimation}, where the authors define a non-parametric estimator for this region using a One-Class Neighbour Machine problem. The idea is to construct a binary classifier that separates a given set of realisations of $C$ between those belonging to the $\mathcal{R}_{1-\alpha}(C)$ and those that do not. The classifier provides the same decision rule $D$ as before, that is $D(\mathbf{c}) = 1$ if $\mathbf{c}$ corresponds to the $1-\alpha$ proportion of elements that belongs to a low entropy (high density) region and zero otherwise. \cite{martos2018entropy} proved that the proposed estimator asymptotically converges to the true $\mathcal{R}_{1-\alpha}(C)$ as the sample size increases.
% ----------------------------------------------------------------------------
\subsection{Bootstrap procedure for FTS}\label{bootstrap}

When constructing prediction intervals or regions, one important issue is the assumption made with respect to the distribution of the innovations of the process. The standard approach is to assume Gaussian innovations, which generate prediction intervals centered on the conditional expectation function, and in general do not constitute a good probabilistic framework when dealing with real time series data. In that context, the bootstrap technique arose as a method that allows for the approximation of the estimator distribution throughout drawing random samples from the empirical distribution function. When dealing with time-dependent data such as functional time series, the usual bootstrap techniques lead to inconsistent statistics. There are several methodologies oriented to tackle this problem and proposed bootstrap techniques for FTS (see \cite{shang2018bootstrap,paparoditis2018sieve, chen2019bootstrap}). 

In this paper, we consider an extension to the functional context of the residual bootstrap methodology proposed by \citet{pascual2004bootstrap} for univariate autoregressive models and extended to the multivariate framework by \citet{fresoli2014bootstrap}. In the FTS context, \citet{franke2019residual} give a formal derivation of this procedure and study the statistical properties of the bootstrapped estimators. They also prove that the method provides asymptotically valid estimates of the mean function and covariance operator as well as the convergence of the empirical distribution of the centered innovations.

Given $Z_1, \ldots, Z_n$ that follow an ARH(1) process as in Eq.~\eqref{eq:arh} the functional residual bootstrap procedure is as follows. 

\begin{enumerate}
\item Obtain the full sample estimators $\hat{\mu}$ and $\hat{\Psi}_K$ using the $n$ realisations;
\item estimate the fitted values $\hat Z_{k}  = \hat \mu + \hat \Psi_K (Z_{k-1} - \hat \mu )$ with an ARH(1) model with $k = 2,\dots,n$;
\item obtain the residuals: $\hat{\epsilon}_k^\dag = \hat{Z}_k - Z_k$ and standardise them.%centre them by subtracting the mean $\hat{\epsilon}_k = \hat{\epsilon}_k^\dag - \bar{\hat{\epsilon}}_k^\dag$;
\item for $b = 1, \ldots B$,
\begin{enumerate}
\item obtain $\hat{\epsilon}_k^*$ by resampling with replacement from $\hat{\epsilon}_k$, and construct the functional bootstrap series $Z_k^* = \hat{\mu}^* + \hat{\Psi}_K (Z_{k-1} - \hat{\mu}^*) + \hat{\epsilon}_k^*$ fixing the first initial curve $Z_1^* = Z_1$;
\item with $Z_1^*, \ldots, Z_n^*$ obtain $\hat{\Psi}_K^*$;
\item obtain the one-step-ahead forecast $\widehat{Z}_{n+1|n}^* = \hat{\mu}^* + \hat{\Psi}_K^* (Z_{n} - \hat{\mu}^*)$, conditioning on the original FTS and store $\widehat{Z}_{n+1|n}^{(b)} := \widehat{Z}_{n+1|n}^*$.
\end{enumerate}
\end{enumerate}

At the end of the procedure, we have $B$ bootstrap predictive pseudo-replicates $\widehat{Z}^* = \{\widehat{Z}_{n+1|n}^{*(1)}, \dots, \widehat{Z}_{n+1|n}^{*(B)}\}$ that follow the predictive law of $Z_n$ as shown in \citet{franke2019residual}.
 
% ----------------------------------------------------------------------------
\subsection{Constructing the predictive confidence bands}
Now we are able to define the simultaneous PCB$_{1-\alpha}$ for a given one-step-ahead functional prediction $\widehat{Z}_{n+1|n}$ as predictive confidence bands with probability $1-\alpha$. At this stage it is important to recall that in the functional context a point prediction refers to the prediction of the whole function. Let $\mathbf{c} = \{\textbf{c}_{d}^{(1)},\dots,\textbf{c}_{d}^{(B)}\} \in \mathbb{R}^d$ be the RKHS multivariate representation of $\widehat{Z}^*$ as defined in Section \ref{ss:rkhs}. With this, we can estimate the MES $\mathcal{\widehat{R}}_{1-\alpha}(\mathbf{c})$ from Eq.~\ref{MES}. Consider the set of indices corresponding to the functions whose multivariate RKHS representation is associated to $\mathcal{\widehat{R}}_{1-\alpha}(\mathbf{c})$, formally $\mathcal{A} = \left\lbrace b \in \lbrace 1, \dots, B\rbrace : \mathbf{c}_d^{(b)} \in \mathcal{\widehat{R}}_{1-\alpha}(\mathbf{c}) \right\rbrace$. Using this, we estimate the predictive confidence band, $\mathcal{B}^{1-\alpha}_{n+1|n}$ for $\widehat{Z}_{n+1|n}$ with the following expression:
\begin{equation}\label{eq:estPCB}
    \hat{\mathcal{B}}^{1-\alpha}_{n+1|n} = \text{Conv}\ \bigg(\bigcup_{b\in\mathcal{A}} G(\widehat{Z}_{n+1|n}^{*(b)})\bigg),
\end{equation}

\noindent
where $G(Z) = \{(t,y): y = Z(t),\ \forall t\in T\}$ is the graph of any function $Z \in \mathcal{H}$ and $\text{Conv}(\cdot)$ refers to the convex hull of the union of the graph of a collection of functions. The functions $\widehat{Z}_{n+1|n}^{*(b)}$ used here, are pseudo-prediction curves obtaining during the $b=1, \ldots, B$ bootstrap iterations. Then, for our PCB, the functional lower bound $L_{n+1|n}$ (respective upper bound $U_{n+1|n}$) equal to the pointwise minimum (respective maximum) of the bootstrap one-step ahead predictors with indices in $\mathcal{A}$. 

The full procedure to construct the predictive confidence band construction %is illustrated in Example \ref{Ex.1} and 
can be summarised as follows:

\begin{enumerate}
    \item Given a raw functional time series data set $Z_k$ and a kernel function $K(s,t)$ we obtain the smoothed curves or functional approximation $\tilde{Z}_k$;
    \vspace{0.25cm}
    \item Estimate the model using Eq.~\ref{eq:arh} and Eq.~\ref{eq:estim-psi} and obtain the fitted values $\hat{Z}_{k}$;
    \vspace{0.25cm}
    \item Following the bootstrap procedure, obtain the bootstrap predictive pseudo-replicates $Z^* = \{Z_{n+h}^{(1)},\dots,Z_{n+h}^{(B)}\}$ and project them onto the same multivariate space used for the estimation of the model in [2];
    \vspace{0.25cm}
    \item Solve the optimization problem in Eq.~\ref{MES} and construct the $\mathcal{\widehat{R}}_{1-\alpha}$ for a desired level of confidence, usually $1-\alpha = \{0.8, 0.9, 0.95\}$;
    \vspace{0.25cm}
    \item Identify the predictive pseudo-replicates whose RKHS representation (the multivariate coefficients $\hat{\mathbf{c}}^* \in \mathbb{R}^d$) belong to the $\mathcal{\widehat{R}}_{1-\alpha}$ and construct the $\text{PCB}_{1-\alpha}$, $\hat{\mathcal{B}}^{1-\alpha}_{n+1|n}$, applying Eq.~\ref{eq:estPCB};
    \item To measure the coverage of the band, check if $G(Z_{n+h}) \in \hat{\mathcal{B}}^{1-\alpha}_{n+1|n}$. This step is only possible when testing data available. 
\end{enumerate}

\section{Experiments}\label{sec:experiments}
The main result of the present work is the construction of a predictive band. This section aims to illustrate and assess the performance of this task. In addition, pointwise prediction is also evaluated. A Monte Carlo study of 500 replicates is performed to assess the quality of our method. We use the pointwise prediction using the root mean squared error (RMSE). To evaluate the predictive band, we construct the simultaneous $\text{PCB}_{1-\alpha}$ for different levels of nominal coverage $1 - \alpha = \{0.8, 0.9, 0.95\}$ and then report the empirical coverage for each case. The average empirical coverage accounts for the average number of times that the function $Z_{n+1}$ is entirely covered by the band simultaneously (that is: for every point in the domain $T$). Moreover, we report the pointwise coverage which shows the percentage of the domain that the function $Z_{n+1}$ is covered by the band. While the empirical coverage gives insightful information, one can construct an arbitrarily good band by sufficiently enlarge its amplitude (see comment below on how this may impact the calibration). Then some notion of efficiency is required to keep the bands as narrow as possible. In this sense, we report the amplitude of the band as a measure of its bandwidth, $\text{Amp} = \int_T (U_{n+h}(t) - L_{n+h}(t))dt.$

\subsection{Simulated Linear Functional Time Series\label{ss:simulated_linear}}

\paragraph{Simulation setting}
To explore the performance of the proposed method in different scenarios, we consider six settings combining different choices for the autoregressive operator $\Psi$ and the covariance operator of the noise $\Gamma_{\epsilon}$. Concerning the operator $\Psi$, we consider a Gaussian operator: $\Psi(t,s) = C(\exp(-(t^2+s^2)/2))$ on one side, and an asymmetric operator $\Psi(t,s) = C(\exp(-s)+\exp(-t+(s^2)/2))$ on the other side. The constant $C$ is chosen such that $\|\Psi\| = 0.5$. We experiment with different covariance operator of the noise, $\Gamma_{\epsilon}$, covering different random sources: normal, Laplace and exponential ones. For their constructions, we develop the operator over a given basis:%, that is we write,
\[ \sum_{k=1}^{K'} \lambda_k \zeta_k \varphi_k(t) \otimes \varphi_k(s),\]
where $\lambda_k \in  \mathbb{R}$, $\zeta_k$ are independent random variables with zero mean and $\varphi_k(t)$ are known functions. 

In our experiments, we set $\varphi_k(t) = \sqrt{2}\sin(\pi (k-1/2) t)$ and draw $\zeta_k$ from different sources. First, we use a standard Gaussian distribution that yields the operator $\Gamma_{\epsilon}$ which is an approximation of the covariance of a Wiener process $\min (t, s)$ if we set $\lambda_k = (\pi (k-1/2))^{-1}$. The other two choices are obtained by modifying the random source only. We draw $\zeta_k$ from a Laplace and an Exponential distribution, respectively. In both cases $E(\zeta_k) = 0$, and we fix the variances to be 2 and 4 for the Laplace and Exponential, respectively.
We introduce these random sources to mimic heavy-tailed and asymmetric noise structures. 
%The Laplace and exponential cases are a way to introduce heavy tails in the distribution of the noise. 
% sigma^2_{noise} for                           | Correction factor
%    normal = 1                                 |     1 (no correction)
%       exp = 4  (var=1/lambda^2)               |    .5 (or lambda = 1)
%   Laplace = 2 (var=2*b^2, where b=1/lambda=1) |    1/sqrt(2) (or lambda = 1/sqrt(2))
%The second data set is simulated according to a non-linear FTS. 
With some language abuse, we refer to them as Wiener, Laplace and Exponential random sources. 
These experiments show that the predictive band remains effective if the prediction method is well-chosen. Finally, the real data set evaluates the efficiency of the proposed approach when the model assumptions are uncontrolled.

\paragraph{Sample sizes} 
We consider five scenarios with different sample sizes generated at $64$ equally spaced time points in the interval $T = [0, 1]$. At each Monte Carlo iteration, we simulated a functional time series of length $1051 (=50+1000+1)$, we drop the first $n_b = 50$ functions (burn-in period) and save the last one for testing purposes. Using the remaining functions, we select the last $N = \{50, 100, 250, 500, 1000\}$ to conform the different sample sizes of our simulation scheme. This way to construct the simulated time series ensures that the test function used in the different scenarios is always the same, which unveils more easily the asymptotic behaviour of the prediction methods.

\paragraph{Competitors}
From a pointwise prediction point of view, we test our proposed methodology (ARH-RKHS) against four prediction methods. As simple baseline approaches, we consider the (historic) mean of the process $\widehat Z_ {n + 1| n} = \hat \mu $ ('Mean') and the predictor by persistence $\widehat Z_ {n + 1|n} = Z_{n}$ ('Persistence'). These naive predictors are associated to two limit cases of the ARH, that is, when the operator norm of $\Psi$ is close to zero and one, respectively. As functional predictors, we use the FAR estimation based on functional principal components ('FAR') as proposed in \citep{bosq2000linear} and implemented in the \texttt{R}-package \texttt{`far'} \citep{julien2015package} and the Functional Partial Least Square Regression ('FPLSR') \citep{delaigle2012methodology} already implemented in the \texttt{R}-package \texttt{`ftsa'} \citep{hyndman2020package}. 
From a predictive band point of view, three alternative methods are used to compare against our entropy procedure of functional predictive confidence bands, hereafter (\texttt{fpcb}). We consider the $L_2$ distance and two depth measures that allow us to construct simultaneous confidence bands for a functional data set: the modified band depth (MBD) \citep{lopez2009concept} and the random projection depth (RPD) \citep{cuevas2007robust}, already implemented in the \texttt{R}-packages \texttt{`depthTools'} \citep{lopez2013depthtools} and \texttt{`fda-usc'} \citep{febrero2013package} respectively. We also construct simultaneous bands using Gaussian Process (GP) as a baseline benchmark.
%Gaussian and empirical pointwise bands which involve applying the usual Gaussian and empirical quantiles to the bootstrap pseudo-replicates $\widehat{Z}^*$ at each point of the domain $t\in T$. 
Our proposal is implemented in the  \texttt{R}-package \texttt{`fpcb'} \citep{fpcb}.

\paragraph{Calibration}
In what follows, we consider the Gaussian kernel function $\allowbreak K(t,s) = \exp({-\sigma (t-s)^2})$ for the estimation of the ARH-RKHS. 
For both the prediction and inferential exercises all the hyperparameters are calibrated through grid search over a predefined temporal window dividing the sample for training and validation (the percentages are specified for each experiment). These hyperparameters are: the bandwidth parameter of the kernel ($\sigma$) and the dimension of the basis function system ($d$) for the ARH-RKHS; the number of functional principal components for the FAR and FPLSR methods. 
To calibrate the hyperparameters of each model, we divide the FTS into the usual training-validation fashion (80\%-20\%) and we simulate an extra curve to test the results. 
The number of bootstrap pseudo-replicates B for the proposed predictive band is set to 1000 and remain fixed for all the numerical experiments. The optimal parameter is the one that minimises a particular metric. In the prediction task this metric is the RMSE and for the task of the $\text{PCB}_{1-\alpha}$ construction the metric is the pinball loss (see \cite{koenker1978regression}).  Given a target $z$, an estimate $\hat{z}$, and quantile level $\tau \in[0,1]$, the pinball loss $\rho_{\tau}$ is defined as
\[ \rho_{\tau}(z, \hat{z})=(\hat{z}-z)(\mathbb{I}\{z \leq \hat{z}\}-\tau).\]
Given the confidence level, say $1-\alpha$, we calibrate our approach to minimize $\rho_{1 - \alpha}$, that is we match the quantile $\tau$ to the desired confidence level.

\paragraph{Results}
We first discuss mean prediction results using Figure 2. It contains box plot for the RMSE of the six prediction scenarios using the five prediction methods.
The ranking of the methods is consist across all the scenarios. The three variants around autoregression (ARH-RKHS, FAR and FPLRS) perform similarly presenting similar mean performance, variability and outliers (in terms of RMSE). Persistence comes into fourth position presenting a slightly higher error level but with similar variability. Finally, the naive method has the worst performance with significant higher error mean level and higher variability. It is worth to mention that while the ranking is consistent, the differences between predictor become more clear when the prediction task is more difficult. In particular,
%The obtained RMSE show that the ARH-RKHS presents a good performance among the functional methods for the different simulation scenarios, although the difference is not statistically significant. Is interesting to mention that 
as we move away from the Gaussian-Wiener setting, the Persistence and Naive prediction methods start performing poorly. In terms of the predictive bands coverage, the results are given in Figure \ref{Simulations} (see also Tables \ref{tab.Sim1}--\ref{tab.Sim6}). The proposed \texttt{fpcb} method shows the good results in terms of empirical coverage throughout all the settings, see Figure \ref{Simulations}. In this sense, the \texttt{fpcb} method is able to reach the target coverage for the different scenarios. In the Gaussian-Wiener setting all the five methods considered but the most efficient ones are the \texttt{fpcb} and the Gaussian Process (as it is expected). In this experiment, the efficiency is measured as the bandwidth (area inside the band). As we move away of the Gaussian-Wiener setting, the GPs start to perform poorly in terms of coverage. In this context of asymmetry of the autocorrelation operator $\Psi$ and non-gaussianity of the covariance of the noise operator $\Gamma_{\epsilon}$ the \texttt{fpcb}, MBD, RPD and L2 present a good performance in terms of coverage but our competitors are less efficient. In the cases where the competitors reach the target coverage, it is at the cost of wider bands; on average 2.6, 3.6 and 2.3 times wider than the \texttt{fpcb} bands for the 80\%, 90\%, 95\% levels of confidence respectively. This means that with the \texttt{fpcb} methodology we can get similar levels of coverage with narrower bands, or have a larger coverage with bands with similar levels of amplitude.

\begin{figure}[htpb]
  \centering
\includegraphics[width=\textwidth]{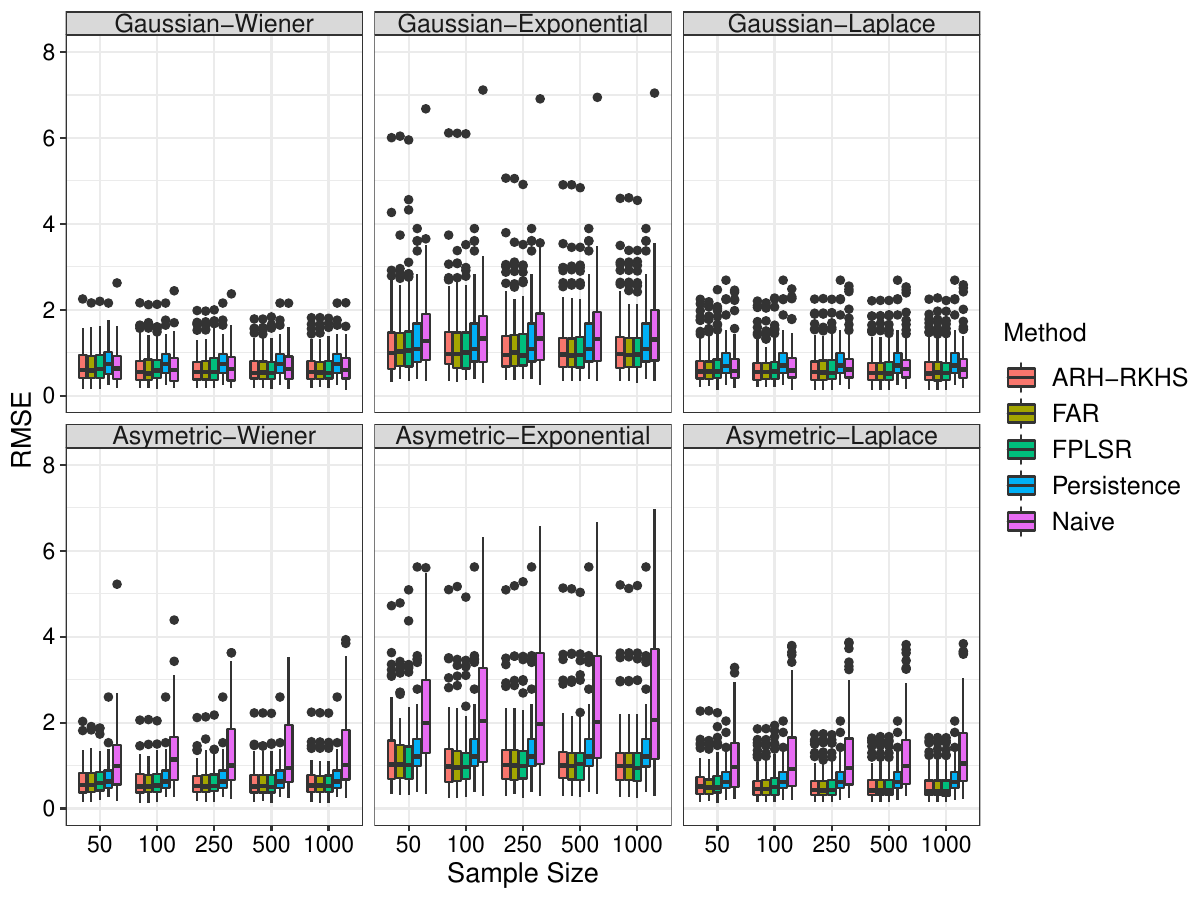}
    \caption{Monte Carlo Results: RMSE distribution by prediction method and simulation setting, throughout a variation of the sample size from 50 to 1000.\label{Fig.RMSE-Sim}}
\end{figure}

\begin{figure}[!ht]
\centering
\caption{Monte Carlo experiment: mean empirical coverage and median amplitude for the different simulations settings, target coverages, band construction methods and sample sizes; 500 MC replications.}
\begin{subfigure}[h]{0.49\textwidth}   
\centering 
\includegraphics[width=\textwidth]{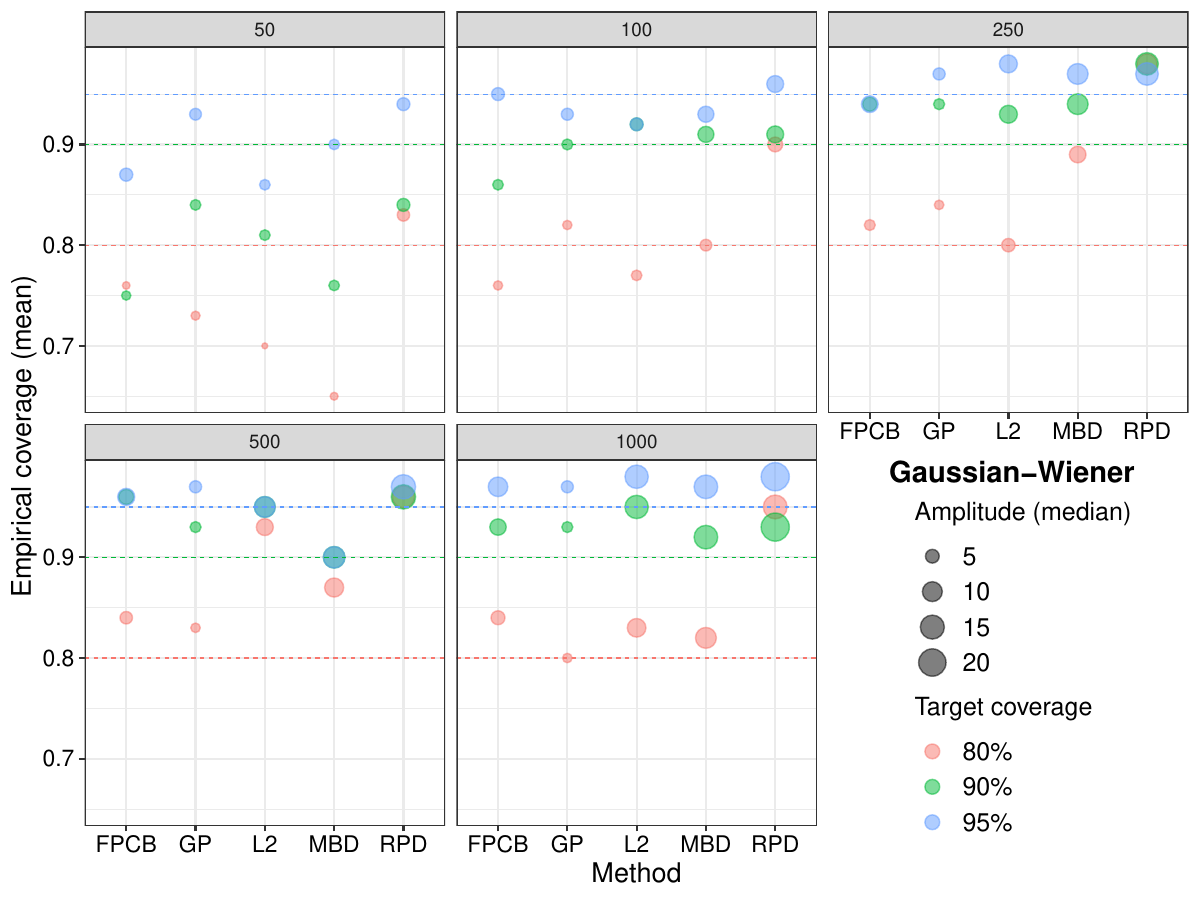}
\end{subfigure}
\hfill
\begin{subfigure}[h]{0.49\textwidth}   
\centering 
\includegraphics[width=\textwidth]{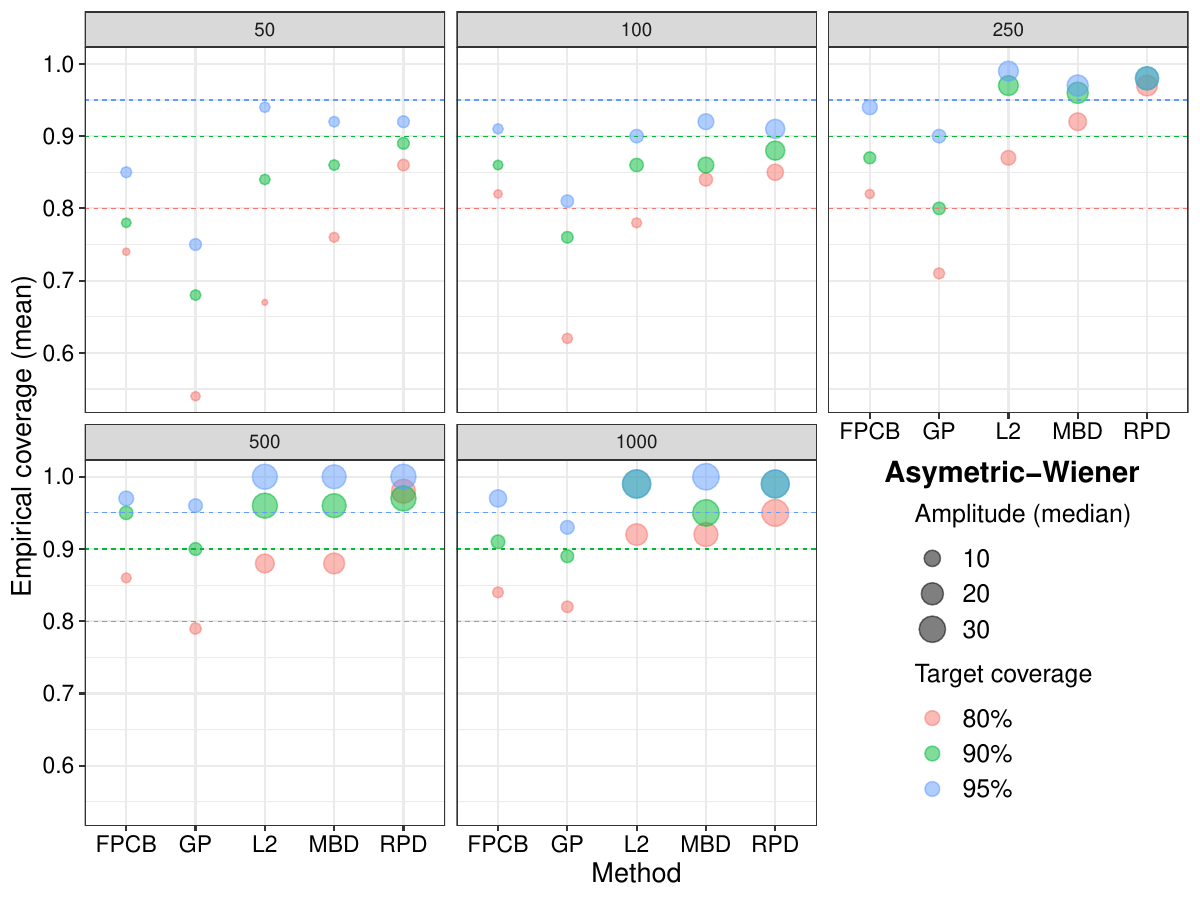}
\end{subfigure}
\hfill
\begin{subfigure}[h]{0.49\textwidth}   
\centering 
\includegraphics[width=\textwidth]{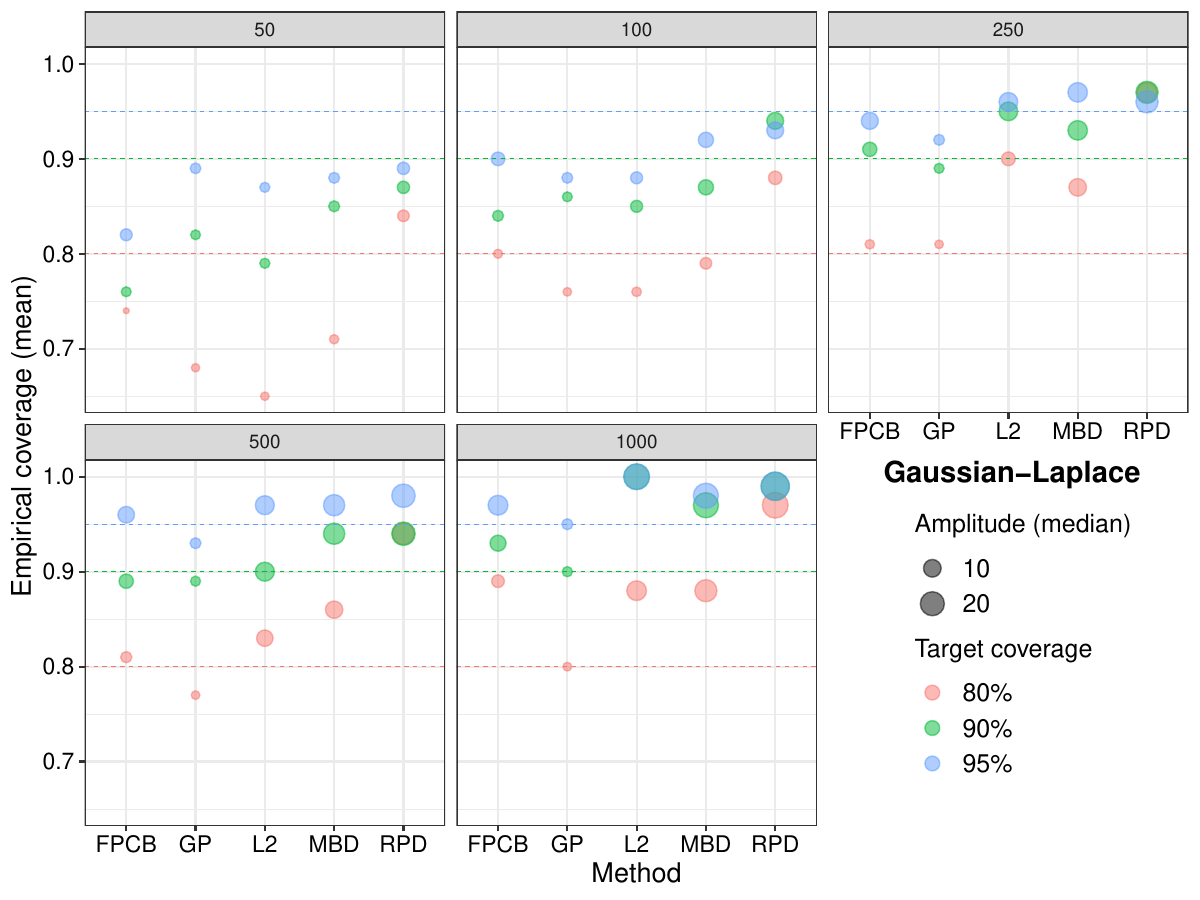}
\end{subfigure}
\hfill
\begin{subfigure}[h]{0.49\textwidth}   
\centering 
\includegraphics[width=\textwidth]{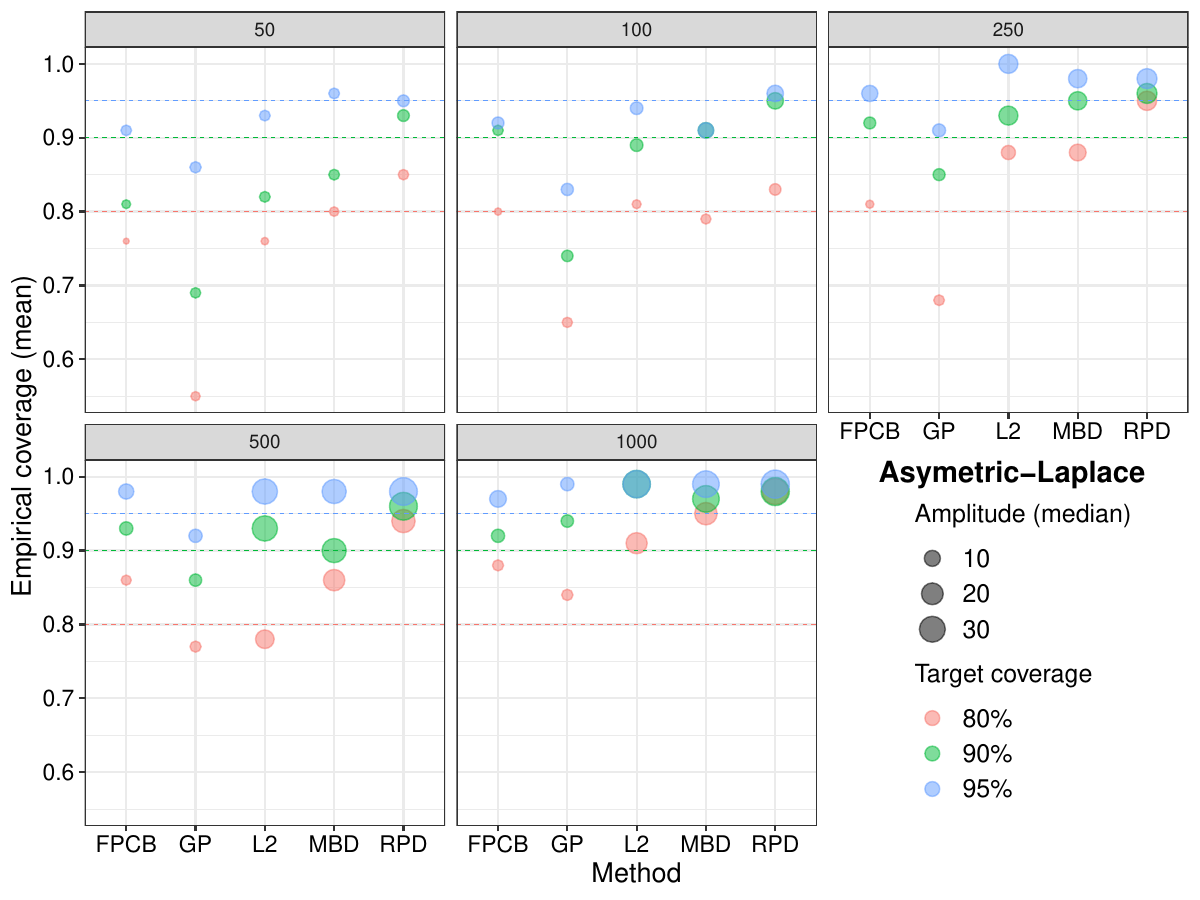}
\end{subfigure}
\hfill
\begin{subfigure}[h]{0.49\textwidth}   
\centering 
\includegraphics[width=\textwidth]{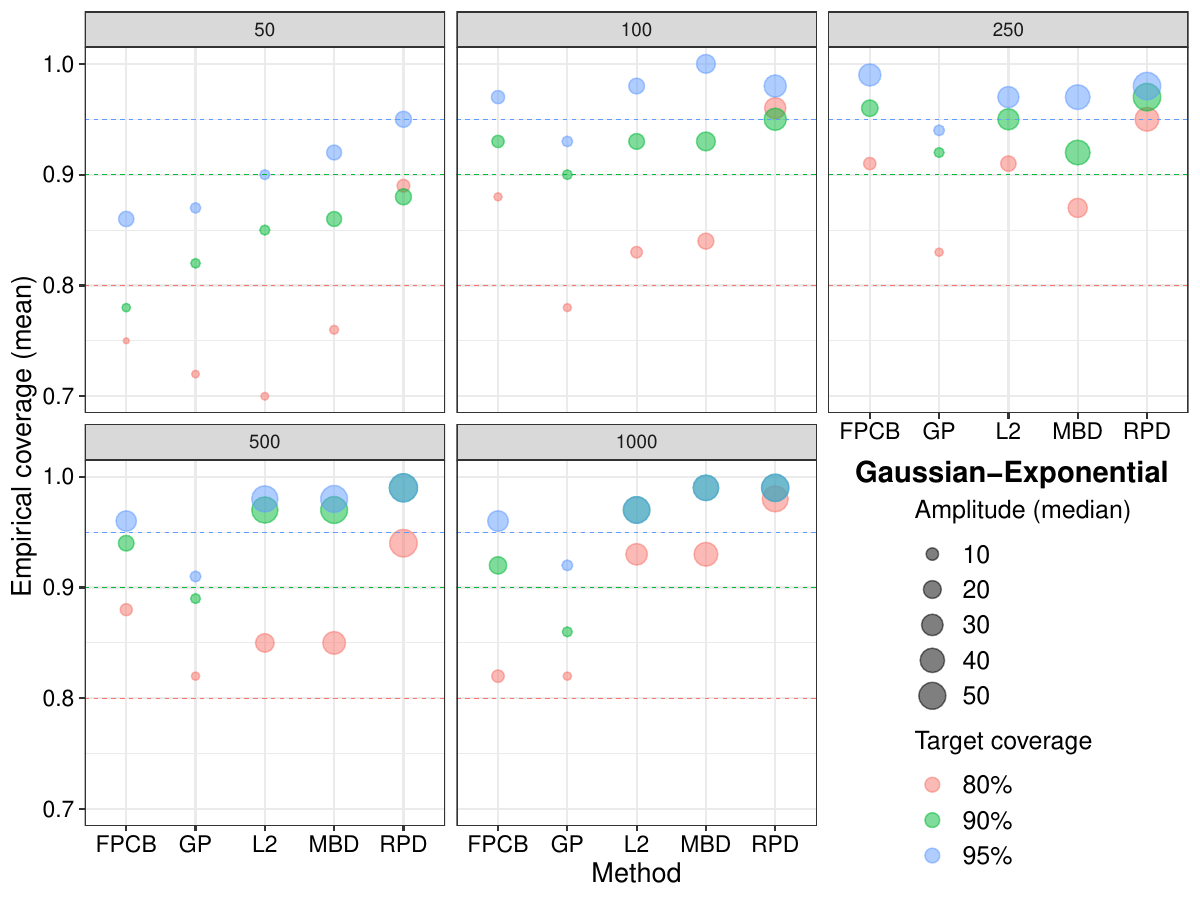}
\end{subfigure}
\hfill
\begin{subfigure}[h]{0.49\textwidth}   
\centering 
\includegraphics[width=\textwidth]{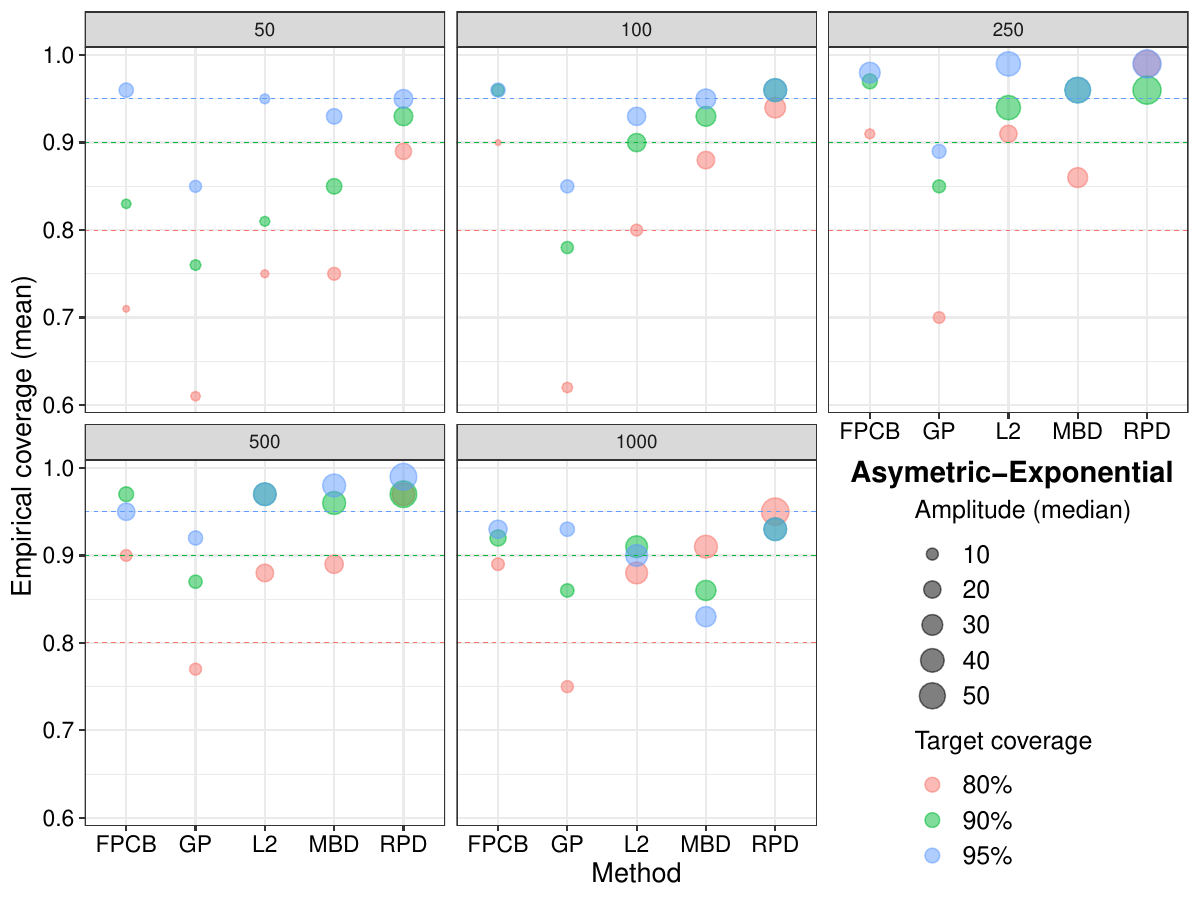}
\end{subfigure}
\label{Simulations}
\end{figure}

\subsection{Experiments on Non-Linear Functional Time Series simulations}

Our contribution mainly addresses the construction of the band once a prediction method was chosen and an appropriate bootstrap scheme was used. In the previous experiment, an ARH model was used for both simulation and prediction, and consequently the results are very good both from pointwise and simultaneous prediction point of views. Now, data are simulated according a non-linear generation process. As expected, in this case the ARH-based prediction model will provide bad prediction, and consequently the confidence band based on this bad prediction will not be too wide. But the combination of the proposed band construction with a prediction model adapted to the data generating process, provides much better results.    

\paragraph{Simulation setting}
For this experiment, data are simulated as in \cite{antoniadis2006functional}, where the authors propose a non-linear functional time series:
\[ X(t)  = \beta_1 m_1(t) + \beta_2 m_2(t) + \epsilon (t),\]
where $m_1 = \cos(2\pi t/64) + \sin(2\pi t/64)$ and  $m_2 = \cos(2\pi t/6) + \sin(2\pi t/6)$, with $\epsilon_t\sim$MA(1) a moving average of length 1. The process is then discretized and ends up in $n$ segments of length 64 each (in total $n\times64$ time points).  We set the following values for the parameters : $\beta_1=0.8, \beta_2=0.18$, the coefficient of the moving average component is 0.8 and the standard deviation of the noise is 0.61. All the parameters were taken from \cite{antoniadis2006functional}, except from the last one which was optimized in order to match the signal-to-noise ratio of this experiment to the one in Section 4.1.

\paragraph{Results}
Table \ref{tab:otros} presents the coverage and amplitude results for the FPCB bands, constructed either with the ARH-RKHS predictor and with the Functional Kernel Wavelet predictor (FKW, \cite{antoniadis2006functional}).
The idea of FKW is to look for patterns in the past data that are similar to the last observed curve. Then, the prediction is a weighed mean of past curves, where weights increase with the similarity between the current and past contexts. 
The predictor is written as 
$$ \widehat{Z}_{n+1} = \sum_{m=2}^{n} w_{n, m-1} Z_{m-1} (t),$$
where $w_{n, m-1}$ measures the similarity between the curves $n$ and $m-1$ for every curve in the past. Note that the weights are computed between the $n$ and $m$-th curves, but the next one (i.e. $m+1$) enters into the predictor. We ran an experiment using 500 MC replicates with the calibration levels obtained before. The nominal coverage levels were still reached, but this time with narrower intervals.

With the ARH-RKHS predictor, the coverage of the bands are, as expected, too high with respect to the nominal levels. Unsurprisingly, this behaviour is obtained given the that the amplitude of the bands is too large. With the FKW predictor, the empirical coverage levels are lower than those obtained with the linear method (ARH-RKHS), although they reach the nominal values. Moreover, these coverage levels are achieved with bands that are on average 7.5 times narrower. This shows that making inferences with the method proposed in this paper can be efficient regardless of the predictor used, as long as it is efficient.  

\begin{table}\centering
\caption{\scriptsize Coverage and amplitude for the FPCB bands constructed using an ARH-RKHS predictor on a non linear functional time series.}
\label{tab:otros}
\scalebox{0.65}{
    \begin{tabular}{lccccccccc}
    \toprule[0.3ex]
     \multicolumn{1}{c}{\multirow{2}[4]{*}{Method}} & \multicolumn{3}{p{13.995em}}{Nominal 80\%} & \multicolumn{3}{p{13.995em}}{Nominal 90\%} & \multicolumn{3}{p{13.995em} }{Nominal 95\%} \\
 & \multicolumn{1}{p{4.665em}}{Cov.} & \multicolumn{1}{p{4.665em}}{Cov. (pointwise)} & \multicolumn{1}{p{4.665em}}{Amp.} & \multicolumn{1}{p{4.665em}}{Cov.} & \multicolumn{1}{p{4.665em}}{Cov. (pointwise)} & \multicolumn{1}{p{4.665em}}{Amp.} & \multicolumn{1}{p{4.665em}}{Cov.} & \multicolumn{1}{p{4.665em}}{Cov. (pointwise)} & \multicolumn{1}{p{4.665em}}{Amp.} \\
    \midrule
    \texttt{fpcb}-ARH-RKHS & 96\%  & 99.86\%  & 7.01  & 98\%  & 99.90\%  & 11.63  & 98.8\%  & 99.90\%  & 19.20 \\
\texttt{fpcb}-FKW   & 89\%  & 99.75\%  & 1.63  & 94\%  & 99.89\%  & 1.68  & 95\%  & 99.90\%  & 1.72 \\
\bottomrule[0.3ex]
\end{tabular}
}%
\end{table}

\subsection{Real data application}\label{ss:PM10}

We illustrate our proposed methodology with the Particulate Matter Concentrations dataset available in \texttt{R}-package \texttt{`ftsa'}. The data consist of 182 daily curves that measure the concentrations (measured in ug/m3) of particular matter with an aerodynamic diameter of less than 10um (PM10) and half-hourly sampled and taken in Graz-Mitte, Austria from October 1, 2010 until March 31, 2011. A square root transformation is applied in order to stabilize the variance.

For this experiment, we divide the sample in the usual training-validation-testing fashion (60\%-20\%-20\%). Given the sample size, we use 110 curves to train, 36 to validate the parameters, and 36 curves for testing. This means we have a forecasting horizon of 36 days, where each forecast is obtained with a one-step-ahead predictor, e.g., the prediction at N+h is conditional to the information at N+h-1. For this experiment, the value of the kernel parameter $\sigma$ is 1 and 7 basis functions were used for all the estimation methods.

\begin{figure}[!ht]
\centering
\caption{PM10 data set: RMSE distribution by prediction method (left); \texttt{fpcb} predictive confidence bands for different confidence levels (right).}
\begin{subfigure}[h]{0.49\textwidth}   
\centering 
\includegraphics[width=\textwidth]{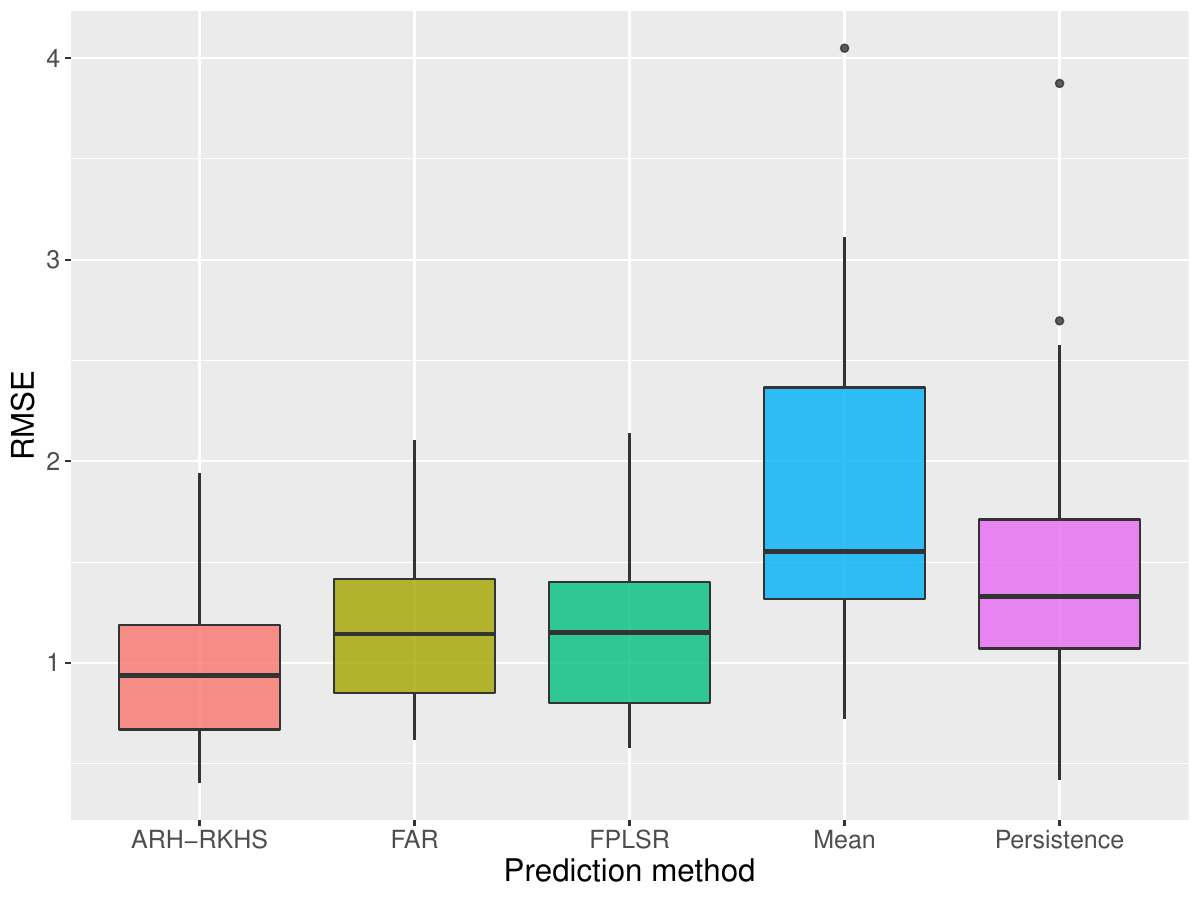}
\end{subfigure}
\hfill
\begin{subfigure}[h]{0.49\textwidth}   
\centering 
\includegraphics[width=\textwidth]{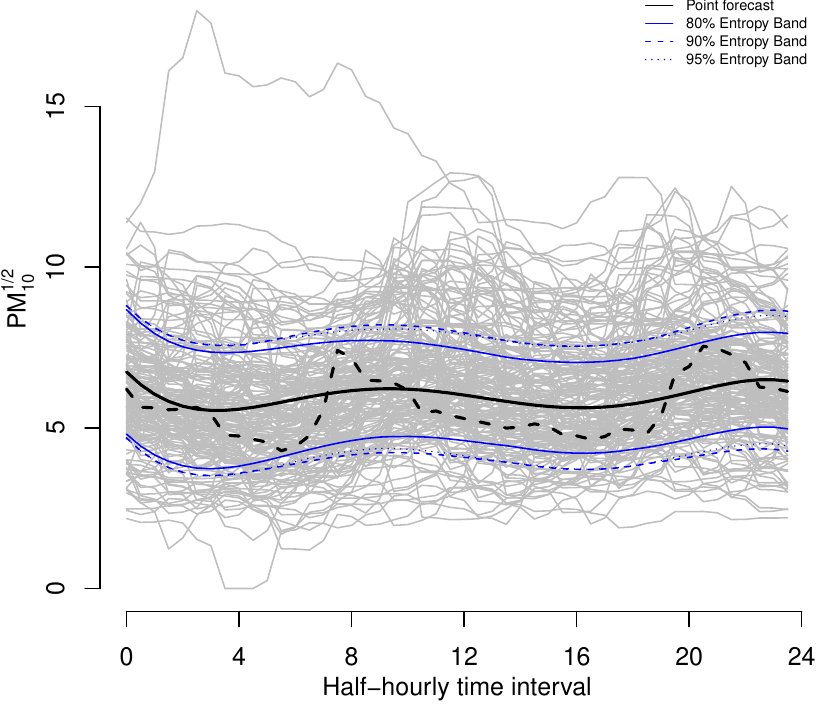}
\label{fig:mean and std of net44}
\end{subfigure}
\label{Fig.pm10}
\end{figure}

The results are presented in Figure \ref{Fig.pm10} and Table \ref{tab.pm10}. In terms of the accuracy of the prediction methods, it can be seen that our proposed model shows a lower RMSE, even if the difference among the functional methods is not statistically significant. However, any of the functional methods improve the quality of the forecast with respect to the persistence and historic mean prediction. The coverage and amplitude results show that the \texttt{fpcb} method offers solid results to make inference. In the three scenarios stated, the \texttt{fpcb} method achieves the desired nominal levels of confidence. This desired confidence level is obtained with wider bands; for a nominal coverage of 95\% the \texttt{fpcb} method is, on average, 30\% wider than the depth-based alternative procedures. The distance-based competitor ($L_2$) presents a good performance in terms of coverage (comparable in some cases with the fpcb method) but with the cost of constructing wider bands. 

Finally, we will comment on the shape of the \texttt{fpcb} bands as illustrated in the right panel of Figure \ref{Fig.pm10}. The target function (black-dashed line) has a form presenting clearly two peaks which are usually associated to traffic rush in the morning and the afternoon. Note that these peaks reflects the behaviour of economic and social patterns and are impacted by the calendar. That is, they may present shifts during the year, caused by time-saving local rules and are more clear during working days than weekends. The prediction has a much smoother shape. Indeed, we aim to estimate the (non-statistical) predictor $\widetilde{Z}_{n+1|n}$ since this is the best we can manage with the information given until moment $n$. Naturally, the band we construct also inherits this constraint, which explains the much smoother behaviour with respect to the datum $Z_{n+1}$.

\begin{table}[htbp]
  \centering
 \caption{\scriptsize PM10 data set: average coverages and amplitude for different nominal coverages ($1-\alpha$), band construction methods and sample sizes; forecast horizon = 36.}
  \renewcommand{\arraystretch}{1} % Tighter table
\scalebox{0.65}{
    \begin{tabular}{lccccccccc}
    \toprule[0.3ex]
    \multicolumn{1}{c}{\multirow{2}[4]{*}{Method}} & \multicolumn{3}{p{15em}}{Nominal 80\%} & \multicolumn{3}{p{15em}}{Nominal 90\%} & \multicolumn{3}{p{15em}}{Nominal 95\%} \\
\cmidrule{2-10}          & \multicolumn{1}{p{5em}}{Cov.} & \multicolumn{1}{p{5em}}{Cov. (pointwise)} & \multicolumn{1}{p{5em}}{Amp.} & \multicolumn{1}{p{5em}}{Cov.} & \multicolumn{1}{p{5em}}{Cov. (pointwise)} & \multicolumn{1}{p{5em}}{Amp.} & \multicolumn{1}{p{5em}}{Cov.} & \multicolumn{1}{p{5em}}{Cov. (pointwise)} & \multicolumn{1}{p{5em}}{Amp.} \\
        \midrule[0.3ex]
    \texttt{fpcb} & 89\%  & 99\%  & 9.17  & 92\%  & 100\% & 9.95  & 95\%  & 100\% & 10.39 \\
    \midrule
    MBD   & 56\%  & 83\%  & 6.21  & 69\%  & 89\%  & 7.49  & 75\%  & 93\%  & 8.30 \\
    \midrule
    RPD   & 58\%  & 84\%  & 6.27  & 67\%  & 91\%  & 7.78  & 78\%  & 94\%  & 8.86 \\
    \midrule
    $L_2$    & 89\%  & 99\%  & 10.63 & 86\%  & 98\%  & 10.61 & 83\%  & 98\%  & 10.66 \\
    \midrule
    $GP$    & 97\%  & 99\%  & 9.85 & 100\%  & 100\%  & 11.28 & 100\%  & 100\%  & 12.68\\
    \bottomrule[0.3ex]
    \end{tabular}%
  }
  \label{tab.pm10}%

\end{table}%

%%%%%%%%%%%%%%%%%%%%%%%%%%%%%%%%%%%%%%%%%%%%%%%%%%%%%%%%%%%%%%%%%%%%%%%%%%%%%%%%%%%%%%%%
%%                                                                                    %%
%%                                  S  E  C  T  I  O  N                               %%
%%                                                                                    %%
%%%%%%%%%%%%%%%%%%%%%%%%%%%%%%%%%%%%%%%%%%%%%%%%%%%%%%%%%%%%%%%%%%%%%%%%%%%%%%%%%%%%%%%%
\section{Concluding comments}
\label{sec:conclusion}

In this paper, we present a novel method for constructing simultaneous confidence bands for the prediction of a functional time series data set, based on an entropy measure for stochastic processes. To construct predictive bands we use a functional version of residual bootstrap that allow us to estimate the prediction law through the use of pseudo-predictions. Each pseudo-realisation is then projected into a space of finite dimension, associated to a functional basis, where the $1-\alpha-$MES are obtained. We map the minimum entropy set back to the functional space and construct a band using the regularity of the RKHS. Through a Monte Carlo study, we show the performance of the proposed model in terms of the accuracy, coverage and amplitude, against other prediction methodologies and band construction techniques for functional time series. To illustrate the procedure through a real data set, an example of particulate matter concentrations is presented. We focus on one-step-ahead predictions for sake of simplicity. Our approach can be applied for horizons $h>1$ if a prediction method and a bootstrap scheme are available for such prediction horizons. Our simulation study demonstrated that the entropy procedure offers a better coverage which improves (reaching in some cases the nominal value) as the sample sizes increases. For both numerical experiments, our proposed method for the band construction achieves a good compromise between coverage and efficiency, the latter measured through the bandwidth.

While two different bootstrap schemes were used, they rely on prediction methods only through some parameters (i.e. $\hat{\mu}$ and $\Psi$ for the ARH-RKHS model; or $w_{n,m}$ for the non-linear predictor). An interesting perspective for 
further work would be to use the approach of \cite{paparoditis2018sieve}, which
relies on the conditional distribution of the one-step ahead estimator $\hat{Z}_{n+1|n}$. 

A key aspect of our method is the construction of the band using a convex hull, as defined in Eq.~\eqref{eq:estPCB}. This implies that the band is compact, leaving aside non-connected configurations. If that were the case, our construction would result in wider bands than the optimal ones. Of course, such situations may arise in practice, for instance if the functional time series presents different regimes or seasonal patterns. Indeed, these non-linear or non-stationary settings need more intrincate constructions that may be investigated in future work.

An important connection with more classical time series can be made. Actually, prediction on FTS can be viewed as simultaneous multi-step-ahead prediction, even in the case where the one-step-ahead function is computed. Take, for instance, the case in section \ref{ss:PM10}. Since each function covers 48 single time points, the FTS approach produces a simultaneous forecast for 48 different horizons. Our approach to computing prediction bands is therefore easily adapted to the construction of prediction intervals for several horizons simultaneously. This is the case for instance in \citet{Staszewska2011} where simultaneous prediction bands are constructed for the trajectories of an impulse-response effect on vector autoregressive processes. Also in \citet{lobo}, the problem is studied from the perspective of joint prediction regions. A comparison of these approaches is given, within the framework of non linear and non stationary FTS in \citet{ANTONIADIS2016939}.

\section*{Declarations}\footnotesize 
\noindent \textbf{Conflicts of interest}: None of the authors has a conflict of interest. \vspace{0.2cm}

\noindent \textbf{Ethics approval}: Authors have no affiliations with or involvement in any organisation or entity with any financial interest or non-financial interest in the subject matter or materials discussed in this manuscript. \vspace{0.2cm}

\noindent \textbf{Consent for publication}: Authors give consent for publication. \vspace{0.2cm}

\noindent \textbf{Availability of data and code}: Data is available in  \texttt{R}-package \texttt{`ftsa'}. and source code to reproduce the results is included as a supplementary file in the submission. \vspace{0.2cm}

\noindent \textbf{Authors’ contributions}: Authors contributed equally to this work.

\appendix
\section{Additional results}
We present here the detailed results for the experiment in Section \ref{ss:simulated_linear}.

\begin{landscape}

\newpage
% Table generated by Excel2LaTeX from sheet 'Gaussian-Wiener'
\begin{table}[htbp]
    \centering
\caption{Monte Carlo experiment. Setting: Gaussian kernel and Wiener noise. Metrics: Mean and standard deviation for simultaneous coverage and point coverage. Mean, median and standard deviation for amplitude. --\scriptsize Results for different target coverages, band construction methods and sample sizes; 500 MC replications.--}
\renewcommand{\arraystretch}{1} % Tighter table
\scalebox{0.7}{   
    \begin{tabular}{ccccccccccccccccc}
    \toprule
    \multicolumn{1}{c}{\multirow{2}[4]{*}{\textbf{Sample Size}}} & \multirow{2}[4]{*}{\textbf{Metrics}} & \multicolumn{3}{p{15em}}{\textbf{Entropy}} & \multicolumn{3}{p{15em}}{\textbf{GPs}} & \multicolumn{3}{p{15em}}{\textbf{MBD}} & \multicolumn{3}{p{15em}}{\textbf{RPD}} & \multicolumn{3}{p{15em}}{\textbf{L2}} \\
\cmidrule{3-17}          &       & \textbf{80\%} & \textbf{90\%} & \textbf{95\%} & \textbf{80\%} & \textbf{90\%} & \textbf{95\%} & \textbf{80\%} & \textbf{90\%} & \textbf{95\%} & \textbf{80\%} & \textbf{90\%} & \textbf{95\%} & \textbf{80\%} & \textbf{90\%} & \textbf{95\%} \\
    \midrule
    \multirow{7}[6]{*}{\textbf{N = 50}} & \multirow{2}[2]{*}{\textbf{Sim. Cov.}} & 0.760 & 0.750 & 0.870 & 0.730 & 0.840 & 0.930 & 0.650 & 0.760 & 0.900 & 0.830 & 0.840 & 0.940 & 0.700 & 0.810 & 0.860 \\
          &       & 0.430 & 0.440 & 0.340 & 0.450 & 0.370 & 0.260 & 0.480 & 0.430 & 0.300 & 0.380 & 0.370 & 0.240 & 0.460 & 0.390 & 0.350 \\
\cmidrule{2-17}          & \multirow{2}[2]{*}{\textbf{Point Cov}} & 0.910 & 0.920 & 0.950 & 0.940 & 0.980 & 0.990 & 0.870 & 0.900 & 0.970 & 0.940 & 0.940 & 0.980 & 0.900 & 0.930 & 0.960 \\
          &       & 0.200 & 0.180 & 0.170 & 0.140 & 0.090 & 0.070 & 0.260 & 0.220 & 0.130 & 0.170 & 0.160 & 0.090 & 0.200 & 0.180 & 0.140 \\
\cmidrule{2-17}          & \multirow{3}[2]{*}{\textbf{Amplitude}} & 3.340 & 3.810 & 5.590 & 3.060 & 3.570 & 3.980 & 3.350 & 4.570 & 4.570 & 5.550 & 5.580 & 5.580 & 3.110 & 3.910 & 3.910 \\
          &       & 2.780 & 3.110 & 4.610 & 3.050 & 3.560 & 3.970 & 2.820 & 3.520 & 3.520 & 4.240 & 4.530 & 4.530 & 2.670 & 3.530 & 3.530 \\
          &       & 1.620 & 1.970 & 4.000 & 0.330 & 0.440 & 0.520 & 1.930 & 4.980 & 4.980 & 6.820 & 6.370 & 6.370 & 1.420 & 1.910 & 1.910 \\
    \midrule
    \multirow{7}[6]{*}{\textbf{N = 100}} & \multirow{2}[2]{*}{\textbf{Sim. Cov.}} & 0.760 & 0.860 & 0.950 & 0.820 & 0.900 & 0.930 & 0.800 & 0.910 & 0.930 & 0.900 & 0.910 & 0.960 & 0.770 & 0.920 & 0.920 \\
          &       & 0.430 & 0.350 & 0.220 & 0.390 & 0.300 & 0.260 & 0.400 & 0.290 & 0.260 & 0.300 & 0.290 & 0.200 & 0.420 & 0.270 & 0.270 \\
\cmidrule{2-17}          & \multirow{2}[2]{*}{\textbf{Point Cov}} & 0.920 & 0.950 & 0.980 & 0.960 & 0.980 & 0.990 & 0.930 & 0.970 & 0.970 & 0.980 & 0.970 & 0.990 & 0.940 & 0.960 & 0.980 \\
          &       & 0.190 & 0.150 & 0.120 & 0.110 & 0.090 & 0.070 & 0.180 & 0.120 & 0.120 & 0.090 & 0.100 & 0.060 & 0.160 & 0.130 & 0.090 \\
\cmidrule{2-17}          & \multirow{3}[2]{*}{\textbf{Amplitude}} & 3.580 & 4.170 & 6.900 & 3.140 & 3.660 & 4.090 & 7.520 & 11.250 & 11.250 & 11.070 & 14.420 & 14.420 & 3.970 & 7.050 & 7.050 \\
          &       & 3.110 & 3.530 & 4.540 & 3.120 & 3.660 & 4.090 & 3.930 & 6.560 & 6.560 & 5.630 & 7.230 & 7.230 & 3.530 & 4.480 & 4.480 \\
          &       & 1.880 & 2.030 & 11.090 & 0.240 & 0.300 & 0.360 & 9.240 & 16.320 & 16.320 & 14.500 & 17.780 & 17.780 & 1.980 & 9.700 & 9.700 \\
    \midrule
    \multirow{7}[6]{*}{\textbf{N = 250}} & \multirow{2}[2]{*}{\textbf{Sim. Cov.}} & 0.820 & 0.940 & 0.940 & 0.840 & 0.940 & 0.970 & 0.890 & 0.940 & 0.970 & 0.980 & 0.980 & 0.970 & 0.800 & 0.930 & 0.980 \\
          &       & 0.390 & 0.240 & 0.240 & 0.370 & 0.240 & 0.170 & 0.310 & 0.240 & 0.170 & 0.140 & 0.140 & 0.170 & 0.400 & 0.260 & 0.140 \\
\cmidrule{2-17}          & \multirow{2}[2]{*}{\textbf{Point Cov}} & 0.940 & 0.980 & 0.980 & 0.970 & 0.990 & 0.990 & 0.950 & 0.980 & 0.990 & 0.990 & 0.990 & 0.990 & 0.900 & 0.970 & 0.990 \\
          &       & 0.170 & 0.100 & 0.100 & 0.100 & 0.070 & 0.070 & 0.160 & 0.090 & 0.060 & 0.100 & 0.060 & 0.060 & 0.270 & 0.150 & 0.050 \\
\cmidrule{2-17}          & \multirow{3}[2]{*}{\textbf{Amplitude}} & 4.100 & 7.000 & 10.300 & 3.180 & 3.720 & 4.160 & 10.930 & 14.760 & 14.760 & 17.680 & 20.720 & 20.720 & 6.030 & 10.210 & 10.210 \\
          &       & 3.660 & 5.130 & 7.030 & 3.170 & 3.680 & 4.130 & 6.960 & 11.020 & 11.020 & 10.180 & 13.120 & 13.120 & 4.780 & 8.130 & 8.130 \\
          &       & 2.400 & 6.830 & 11.900 & 0.190 & 0.240 & 0.270 & 11.180 & 13.290 & 13.290 & 23.040 & 24.240 & 24.240 & 4.600 & 7.460 & 7.460 \\
    \midrule
    \multirow{7}[6]{*}{\textbf{N = 500}} & \multirow{2}[2]{*}{\textbf{Sim. Cov.}} & 0.840 & 0.960 & 0.960 & 0.830 & 0.930 & 0.970 & 0.870 & 0.900 & 0.900 & 0.960 & 0.960 & 0.970 & 0.930 & 0.950 & 0.950 \\
          &       & 0.370 & 0.200 & 0.200 & 0.380 & 0.260 & 0.170 & 0.340 & 0.300 & 0.300 & 0.200 & 0.200 & 0.170 & 0.260 & 0.220 & 0.220 \\
\cmidrule{2-17}          & \multirow{2}[2]{*}{\textbf{Point Cov}} & 0.950 & 0.980 & 0.980 & 0.980 & 0.990 & 1.000 & 0.950 & 0.960 & 0.970 & 0.980 & 0.980 & 0.990 & 0.970 & 0.980 & 0.980 \\
          &       & 0.150 & 0.110 & 0.110 & 0.080 & 0.060 & 0.030 & 0.160 & 0.150 & 0.130 & 0.120 & 0.120 & 0.070 & 0.150 & 0.120 & 0.110 \\
\cmidrule{2-17}          & \multirow{3}[2]{*}{\textbf{Amplitude}} & 5.370 & 7.640 & 13.470 & 3.170 & 3.690 & 4.160 & 17.230 & 23.110 & 23.110 & 23.200 & 28.430 & 28.430 & 12.770 & 19.310 & 19.310 \\
          &       & 4.310 & 5.720 & 7.340 & 3.170 & 3.680 & 4.150 & 9.160 & 11.920 & 11.920 & 12.570 & 15.450 & 15.450 & 7.250 & 11.050 & 11.050 \\
          &       & 4.190 & 6.100 & 22.450 & 0.160 & 0.210 & 0.250 & 27.180 & 31.950 & 31.950 & 36.820 & 38.530 & 38.530 & 20.840 & 28.460 & 28.460 \\
    \midrule
    \multirow{7}[6]{*}{\textbf{N = 1000}} & \multirow{2}[2]{*}{\textbf{Sim. Cov.}} & 0.840 & 0.930 & 0.970 & 0.800 & 0.930 & 0.970 & 0.820 & 0.920 & 0.970 & 0.950 & 0.930 & 0.980 & 0.830 & 0.950 & 0.980 \\
          &       & 0.370 & 0.260 & 0.170 & 0.400 & 0.260 & 0.170 & 0.390 & 0.270 & 0.170 & 0.220 & 0.260 & 0.140 & 0.380 & 0.220 & 0.140 \\
\cmidrule{2-17}          & \multirow{2}[2]{*}{\textbf{Point Cov}} & 0.960 & 0.980 & 0.990 & 0.970 & 0.990 & 1.000 & 0.930 & 0.970 & 0.990 & 0.980 & 0.990 & 1.000 & 0.920 & 0.980 & 1.000 \\
          &       & 0.120 & 0.070 & 0.030 & 0.090 & 0.060 & 0.030 & 0.190 & 0.110 & 0.080 & 0.090 & 0.060 & 0.040 & 0.230 & 0.110 & 0.030 \\
\cmidrule{2-17}          & \multirow{3}[2]{*}{\textbf{Amplitude}} & 5.540 & 10.440 & 14.380 & 3.180 & 3.690 & 4.130 & 21.680 & 26.530 & 26.530 & 27.960 & 31.830 & 31.830 & 13.380 & 27.830 & 27.830 \\
          &       & 5.120 & 6.810 & 9.700 & 3.170 & 3.670 & 4.100 & 11.100 & 14.520 & 14.520 & 14.720 & 21.270 & 21.270 & 8.740 & 14.080 & 14.080 \\
          &       & 3.330 & 10.500 & 11.970 & 0.110 & 0.160 & 0.220 & 25.990 & 26.460 & 26.460 & 29.680 & 29.650 & 29.650 & 14.280 & 33.280 & 33.280 \\
\bottomrule[0.3ex]
\end{tabular}%
}
\label{tab.Sim2}%
\end{table}%

\newpage
% Table generated by Excel2LaTeX from sheet 'Gaussian-Wiener'
\begin{table}[htbp]
    \centering
  \caption{Monte Carlo experiment. Setting: Gaussian kernel and Laplacian noise. Metrics: Mean and standard deviation for simultaneous coverage and point coverage. Mean, median and standard deviation for amplitude. --\scriptsize Results for different target coverages, band construction methods and sample sizes; 500 MC replications.--}
\renewcommand{\arraystretch}{1} % Tighter table
\scalebox{0.7}{   
    \begin{tabular}{ccccccccccccccccc}
    \toprule
    \multicolumn{1}{c}{\multirow{2}[4]{*}{\textbf{Sample Size}}} & \multirow{2}[4]{*}{\textbf{Metrics}} & \multicolumn{3}{p{15em}}{\textbf{Entropy}} & \multicolumn{3}{p{15em}}{\textbf{GPs}} & \multicolumn{3}{p{15em}}{\textbf{MBD}} & \multicolumn{3}{p{15em}}{\textbf{RPD}} & \multicolumn{3}{p{15em}}{\textbf{L2}} \\
\cmidrule{3-17}          &       & \textbf{80\%} & \textbf{90\%} & \textbf{95\%} & \textbf{80\%} & \textbf{90\%} & \textbf{95\%} & \textbf{80\%} & \textbf{90\%} & \textbf{95\%} & \textbf{80\%} & \textbf{90\%} & \textbf{95\%} & \textbf{80\%} & \textbf{90\%} & \textbf{95\%} \\
    \midrule
    \multirow{7}[6]{*}{\textbf{N -= 50}} & \multirow{2}[2]{*}{\textbf{Sim. Cov.}} & 0.740 & 0.760 & 0.820 & 0.680 & 0.820 & 0.890 & 0.710 & 0.850 & 0.880 & 0.840 & 0.870 & 0.890 & 0.650 & 0.790 & 0.870 \\
          &       & 0.440 & 0.430 & 0.390 & 0.470 & 0.390 & 0.310 & 0.460 & 0.360 & 0.330 & 0.370 & 0.340 & 0.310 & 0.480 & 0.410 & 0.340 \\
\cmidrule{2-17}          & \multirow{2}[2]{*}{\textbf{Point Cov}} & 0.910 & 0.930 & 0.940 & 0.930 & 0.970 & 0.990 & 0.920 & 0.940 & 0.950 & 0.950 & 0.960 & 0.970 & 0.900 & 0.930 & 0.950 \\
          &       & 0.210 & 0.170 & 0.160 & 0.160 & 0.100 & 0.050 & 0.180 & 0.170 & 0.160 & 0.150 & 0.120 & 0.110 & 0.190 & 0.190 & 0.160 \\
\cmidrule{2-17}          & \multirow{3}[2]{*}{\textbf{Amplitude}} & 3.320 & 4.100 & 8.960 & 3.130 & 3.620 & 4.050 & 4.240 & 6.370 & 6.370 & 6.140 & 8.900 & 8.900 & 3.250 & 4.290 & 4.290 \\
          &       & 2.740 & 3.600 & 4.650 & 3.070 & 3.570 & 3.970 & 3.370 & 4.140 & 4.140 & 4.410 & 4.930 & 4.930 & 3.100 & 3.680 & 3.680 \\
          &       & 1.900 & 2.360 & 20.210 & 0.470 & 0.570 & 0.660 & 3.590 & 9.510 & 9.510 & 7.830 & 13.540 & 13.540 & 1.490 & 3.340 & 3.340 \\
    \midrule
    \multirow{7}[6]{*}{\textbf{N = 100}} & \multirow{2}[2]{*}{\textbf{Sim. Cov.}} & 0.800 & 0.840 & 0.900 & 0.760 & 0.860 & 0.880 & 0.790 & 0.870 & 0.920 & 0.880 & 0.940 & 0.930 & 0.760 & 0.850 & 0.880 \\
          &       & 0.400 & 0.370 & 0.300 & 0.430 & 0.350 & 0.330 & 0.410 & 0.340 & 0.270 & 0.330 & 0.240 & 0.260 & 0.430 & 0.360 & 0.330 \\
\cmidrule{2-17}          & \multirow{2}[2]{*}{\textbf{Point Cov}} & 0.930 & 0.950 & 0.970 & 0.930 & 0.950 & 0.970 & 0.940 & 0.960 & 0.960 & 0.960 & 0.980 & 0.980 & 0.920 & 0.960 & 0.960 \\
          &       & 0.180 & 0.160 & 0.130 & 0.190 & 0.160 & 0.130 & 0.170 & 0.150 & 0.150 & 0.130 & 0.080 & 0.100 & 0.190 & 0.120 & 0.140 \\
\cmidrule{2-17}          & \multirow{3}[2]{*}{\textbf{Amplitude}} & 3.740 & 5.940 & 8.120 & 3.120 & 3.620 & 4.100 & 7.720 & 12.010 & 12.010 & 10.580 & 16.570 & 16.570 & 4.330 & 7.720 & 7.720 \\
          &       & 3.300 & 4.170 & 5.870 & 3.130 & 3.600 & 4.100 & 4.460 & 7.330 & 7.330 & 5.720 & 9.080 & 9.080 & 3.470 & 4.720 & 4.720 \\
          &       & 2.160 & 4.950 & 11.470 & 0.300 & 0.380 & 0.460 & 11.640 & 16.710 & 16.710 & 15.770 & 24.770 & 24.770 & 2.880 & 10.480 & 10.480 \\
    \midrule
    \multirow{7}[6]{*}{\textbf{N = 250}} & \multirow{2}[2]{*}{\textbf{Sim. Cov.}} & 0.810 & 0.890 & 0.940 & 0.810 & 0.890 & 0.920 & 0.870 & 0.930 & 0.970 & 0.970 & 0.970 & 0.960 & 0.900 & 0.950 & 0.960 \\
          &       & 0.390 & 0.290 & 0.240 & 0.390 & 0.310 & 0.270 & 0.340 & 0.260 & 0.170 & 0.170 & 0.170 & 0.200 & 0.300 & 0.220 & 0.200 \\
\cmidrule{2-17}          & \multirow{2}[2]{*}{\textbf{Point Cov}} & 0.920 & 0.970 & 0.970 & 0.950 & 0.960 & 0.970 & 0.970 & 0.970 & 0.990 & 1.000 & 0.990 & 0.990 & 0.950 & 0.980 & 0.990 \\
          &       & 0.210 & 0.120 & 0.120 & 0.170 & 0.160 & 0.140 & 0.110 & 0.140 & 0.080 & 0.020 & 0.040 & 0.040 & 0.170 & 0.130 & 0.050 \\
\cmidrule{2-17}          & \multirow{3}[2]{*}{\textbf{Amplitude}} & 4.380 & 7.600 & 15.600 & 3.160 & 3.680 & 4.120 & 17.880 & 23.650 & 23.650 & 23.100 & 30.530 & 30.530 & 7.660 & 17.930 & 17.930 \\
          &       & 3.430 & 6.450 & 9.290 & 3.140 & 3.660 & 4.090 & 9.860 & 12.430 & 12.430 & 13.160 & 16.660 & 16.660 & 5.970 & 11.330 & 11.330 \\
          &       & 3.090 & 6.310 & 20.060 & 0.230 & 0.270 & 0.340 & 31.680 & 39.390 & 39.390 & 37.740 & 47.750 & 47.750 & 5.490 & 28.470 & 28.470 \\
    \midrule
    \multirow{7}[6]{*}{\textbf{N = 500}} & \multirow{2}[2]{*}{\textbf{Sim. Cov.}} & 0.810 & 0.900 & 0.960 & 0.770 & 0.890 & 0.930 & 0.860 & 0.940 & 0.970 & 0.940 & 0.940 & 0.980 & 0.830 & 0.900 & 0.970 \\
          &       & 0.390 & 0.310 & 0.200 & 0.420 & 0.310 & 0.260 & 0.350 & 0.240 & 0.170 & 0.240 & 0.240 & 0.140 & 0.380 & 0.300 & 0.170 \\
\cmidrule{2-17}          & \multirow{2}[2]{*}{\textbf{Point Cov}} & 0.940 & 0.970 & 0.980 & 0.940 & 0.970 & 0.980 & 0.960 & 0.970 & 0.990 & 0.980 & 0.980 & 0.990 & 0.920 & 0.950 & 0.990 \\
          &       & 0.160 & 0.130 & 0.090 & 0.170 & 0.120 & 0.100 & 0.140 & 0.120 & 0.080 & 0.090 & 0.080 & 0.070 & 0.210 & 0.160 & 0.070 \\
\cmidrule{2-17}          & \multirow{3}[2]{*}{\textbf{Amplitude}} & 4.810 & 8.410 & 15.070 & 3.170 & 3.670 & 4.130 & 16.150 & 23.350 & 23.350 & 22.870 & 29.390 & 29.390 & 11.050 & 18.800 & 18.800 \\
          &       & 4.260 & 6.490 & 8.700 & 3.140 & 3.650 & 4.090 & 9.630 & 15.130 & 15.130 & 16.270 & 18.820 & 18.820 & 8.510 & 11.730 & 11.730 \\
          &       & 2.860 & 7.620 & 18.280 & 0.170 & 0.220 & 0.260 & 19.840 & 23.750 & 23.750 & 24.610 & 30.220 & 30.220 & 10.530 & 22.130 & 22.130 \\
    \midrule
    \multirow{7}[6]{*}{\textbf{N = 1000}} & \multirow{2}[2]{*}{\textbf{Sim. Cov.}} & 0.890 & 0.930 & 0.970 & 0.800 & 0.900 & 0.950 & 0.880 & 0.970 & 0.980 & 0.970 & 0.990 & 0.990 & 0.880 & 1.000 & 1.000 \\
          &       & 0.310 & 0.260 & 0.170 & 0.400 & 0.300 & 0.220 & 0.330 & 0.170 & 0.140 & 0.170 & 0.100 & 0.100 & 0.330 & 0.000 & 0.000 \\
\cmidrule{2-17}          & \multirow{2}[2]{*}{\textbf{Point Cov}} & 0.940 & 0.980 & 0.990 & 0.940 & 0.960 & 0.970 & 0.960 & 1.000 & 1.000 & 0.990 & 1.000 & 1.000 & 0.940 & 1.000 & 1.000 \\
          &       & 0.180 & 0.100 & 0.040 & 0.190 & 0.160 & 0.140 & 0.130 & 0.030 & 0.020 & 0.070 & 0.010 & 0.020 & 0.200 & 0.000 & 0.000 \\
\cmidrule{2-17}          & \multirow{3}[2]{*}{\textbf{Amplitude}} & 5.990 & 14.110 & 20.930 & 3.180 & 3.710 & 4.150 & 23.950 & 32.560 & 32.560 & 34.350 & 39.550 & 39.550 & 22.260 & 32.960 & 32.960 \\
          &       & 5.170 & 8.140 & 12.940 & 3.190 & 3.730 & 4.130 & 16.510 & 22.060 & 22.060 & 23.420 & 28.620 & 28.620 & 12.690 & 23.150 & 23.150 \\
          &       & 4.080 & 16.390 & 23.490 & 0.160 & 0.200 & 0.240 & 22.180 & 29.630 & 29.630 & 33.810 & 37.110 & 37.110 & 26.370 & 29.640 & 29.640 \\
\bottomrule[0.3ex]
\end{tabular}%
}
\label{tab.Sim2}%
\end{table}%

\newpage
% Table generated by Excel2LaTeX from sheet 'Gaussian-Wiener'
\begin{table}[htbp]
    \centering
  \caption{Monte Carlo experiment. Setting: Gaussian kernel and exponential noise. Metrics: Mean and standard deviation for simultaneous coverage and point coverage. Mean, median and standard deviation for amplitude. --\scriptsize Results for different target coverages, band construction methods and sample sizes; 500 MC replications.--}
\renewcommand{\arraystretch}{1} % Tighter table
\scalebox{0.7}{   
    \begin{tabular}{ccccccccccccccccc}
    \toprule
    \multicolumn{1}{c}{\multirow{2}[4]{*}{\textbf{Sample Size}}} & \multirow{2}[4]{*}{\textbf{Metrics}} & \multicolumn{3}{p{15em}}{\textbf{Entropy}} & \multicolumn{3}{p{15em}}{\textbf{GPs}} & \multicolumn{3}{p{15em}}{\textbf{MBD}} & \multicolumn{3}{p{15em}}{\textbf{RPD}} & \multicolumn{3}{p{15em}}{\textbf{L2}} \\
\cmidrule{3-17}          &       & \textbf{80\%} & \textbf{90\%} & \textbf{95\%} & \textbf{80\%} & \textbf{90\%} & \textbf{95\%} & \textbf{80\%} & \textbf{90\%} & \textbf{95\%} & \textbf{80\%} & \textbf{90\%} & \textbf{95\%} & \textbf{80\%} & \textbf{90\%} & \textbf{95\%} \\
    \midrule
    \multirow{7}[6]{*}{\textbf{N = 50}} & \multirow{2}[2]{*}{\textbf{Sim. Cov.}} & 0.750 & 0.780 & 0.860 & 0.720 & 0.820 & 0.870 & 0.760 & 0.860 & 0.920 & 0.890 & 0.880 & 0.950 & 0.700 & 0.850 & 0.900 \\
          &       & 0.440 & 0.420 & 0.350 & 0.450 & 0.390 & 0.340 & 0.430 & 0.350 & 0.270 & 0.310 & 0.330 & 0.220 & 0.460 & 0.360 & 0.300 \\
\cmidrule{2-17}          & \multirow{2}[2]{*}{\textbf{Point Cov}} & 0.920 & 0.930 & 0.940 & 0.930 & 0.960 & 0.970 & 0.930 & 0.960 & 0.970 & 0.980 & 0.970 & 0.980 & 0.910 & 0.960 & 0.970 \\
          &       & 0.190 & 0.170 & 0.170 & 0.170 & 0.130 & 0.120 & 0.160 & 0.110 & 0.130 & 0.070 & 0.110 & 0.110 & 0.200 & 0.130 & 0.110 \\
\cmidrule{2-17}          & \multirow{3}[2]{*}{\textbf{Amplitude}} & 12.030 & 17.140 & 47.030 & 6.070 & 7.090 & 7.940 & 25.860 & 37.030 & 37.030 & 38.970 & 49.080 & 49.080 & 10.240 & 18.890 & 18.890 \\
          &       & 5.740 & 6.440 & 14.680 & 6.050 & 7.080 & 7.850 & 6.710 & 13.990 & 13.990 & 10.440 & 16.150 & 16.150 & 6.110 & 7.450 & 7.450 \\
          &       & 34.500 & 38.240 & 90.740 & 0.910 & 1.170 & 1.410 & 54.450 & 63.660 & 63.660 & 76.450 & 85.830 & 85.830 & 31.110 & 43.240 & 43.240 \\
    \midrule
    \multirow{7}[6]{*}{\textbf{N = 100}} & \multirow{2}[2]{*}{\textbf{Sim. Cov.}} & 0.880 & 0.930 & 0.970 & 0.780 & 0.900 & 0.930 & 0.840 & 0.930 & 1.000 & 0.960 & 0.950 & 0.980 & 0.830 & 0.930 & 0.980 \\
          &       & 0.330 & 0.260 & 0.170 & 0.420 & 0.300 & 0.260 & 0.370 & 0.260 & 0.000 & 0.200 & 0.220 & 0.140 & 0.380 & 0.260 & 0.140 \\
\cmidrule{2-17}          & \multirow{2}[2]{*}{\textbf{Point Cov}} & 0.970 & 0.980 & 0.990 & 0.960 & 0.980 & 0.980 & 0.950 & 0.980 & 1.000 & 0.980 & 0.990 & 1.000 & 0.930 & 0.980 & 0.990 \\
          &       & 0.120 & 0.110 & 0.040 & 0.140 & 0.100 & 0.070 & 0.170 & 0.090 & 0.000 & 0.090 & 0.060 & 0.030 & 0.200 & 0.120 & 0.090 \\
\cmidrule{2-17}          & \multirow{3}[2]{*}{\textbf{Amplitude}} & 9.940 & 20.680 & 27.600 & 6.260 & 7.230 & 8.100 & 32.720 & 45.080 & 45.080 & 55.520 & 55.620 & 55.620 & 13.520 & 28.710 & 28.710 \\
          &       & 6.270 & 9.850 & 11.190 & 6.240 & 7.200 & 8.040 & 15.930 & 22.500 & 22.500 & 29.420 & 32.330 & 32.330 & 8.900 & 15.530 & 15.530 \\
          &       & 11.290 & 39.080 & 43.410 & 0.680 & 0.850 & 0.950 & 45.450 & 60.280 & 60.280 & 75.620 & 66.320 & 66.320 & 16.960 & 39.940 & 39.940 \\
    \midrule
    \multirow{7}[6]{*}{\textbf{N = 250}} & \multirow{2}[2]{*}{\textbf{Sim. Cov.}} & 0.910 & 0.960 & 0.990 & 0.830 & 0.920 & 0.940 & 0.870 & 0.920 & 0.970 & 0.950 & 0.970 & 0.980 & 0.910 & 0.950 & 0.970 \\
          &       & 0.290 & 0.200 & 0.100 & 0.380 & 0.270 & 0.240 & 0.340 & 0.270 & 0.170 & 0.220 & 0.170 & 0.140 & 0.290 & 0.220 & 0.170 \\
\cmidrule{2-17}          & \multirow{2}[2]{*}{\textbf{Point Cov}} & 0.970 & 0.980 & 1.000 & 0.960 & 0.970 & 0.980 & 0.960 & 0.970 & 0.990 & 0.990 & 0.990 & 0.990 & 0.970 & 0.990 & 0.990 \\
          &       & 0.120 & 0.110 & 0.050 & 0.160 & 0.150 & 0.140 & 0.140 & 0.120 & 0.080 & 0.060 & 0.040 & 0.050 & 0.140 & 0.040 & 0.040 \\
\cmidrule{2-17}          & \multirow{3}[2]{*}{\textbf{Amplitude}} & 12.680 & 28.230 & 47.180 & 6.350 & 7.390 & 8.210 & 43.650 & 66.130 & 66.130 & 65.620 & 81.350 & 81.350 & 30.350 & 53.880 & 53.880 \\
          &       & 9.860 & 16.980 & 32.120 & 6.360 & 7.390 & 8.190 & 23.440 & 41.270 & 41.270 & 38.040 & 52.180 & 52.180 & 14.810 & 29.130 & 29.130 \\
          &       & 9.710 & 36.710 & 46.630 & 0.470 & 0.630 & 0.710 & 57.930 & 77.480 & 77.480 & 78.150 & 95.340 & 95.340 & 54.330 & 76.100 & 76.100 \\
    \midrule
    \multirow{7}[6]{*}{\textbf{N = 500}} & \multirow{2}[2]{*}{\textbf{Sim. Cov.}} & 0.880 & 0.940 & 0.960 & 0.820 & 0.890 & 0.910 & 0.850 & 0.970 & 0.980 & 0.940 & 0.990 & 0.990 & 0.850 & 0.970 & 0.980 \\
          &       & 0.330 & 0.240 & 0.200 & 0.390 & 0.310 & 0.290 & 0.360 & 0.170 & 0.140 & 0.240 & 0.100 & 0.100 & 0.360 & 0.170 & 0.140 \\
\cmidrule{2-17}          & \multirow{2}[2]{*}{\textbf{Point Cov}} & 0.950 & 0.980 & 0.980 & 0.960 & 0.980 & 0.980 & 0.950 & 0.990 & 0.990 & 0.980 & 1.000 & 1.000 & 0.940 & 1.000 & 0.990 \\
          &       & 0.150 & 0.100 & 0.100 & 0.120 & 0.090 & 0.070 & 0.160 & 0.060 & 0.060 & 0.130 & 0.020 & 0.050 & 0.190 & 0.020 & 0.070 \\
\cmidrule{2-17}          & \multirow{3}[2]{*}{\textbf{Amplitude}} & 17.640 & 30.670 & 45.920 & 6.340 & 7.350 & 8.200 & 50.620 & 65.860 & 65.860 & 71.090 & 76.820 & 76.820 & 38.650 & 68.920 & 68.920 \\
          &       & 9.520 & 15.290 & 26.940 & 6.330 & 7.320 & 8.210 & 34.260 & 49.520 & 49.520 & 52.690 & 55.320 & 55.320 & 21.780 & 47.190 & 47.190 \\
          &       & 28.170 & 40.000 & 48.550 & 0.380 & 0.450 & 0.540 & 45.890 & 51.750 & 51.750 & 60.120 & 62.330 & 62.330 & 42.530 & 77.280 & 77.280 \\
    \midrule
    \multirow{7}[6]{*}{\textbf{N = 1000}} & \multirow{2}[2]{*}{\textbf{Sim. Cov.}} & 0.820 & 0.920 & 0.960 & 0.820 & 0.860 & 0.920 & 0.930 & 0.990 & 0.990 & 0.980 & 0.990 & 0.990 & 0.930 & 0.970 & 0.970 \\
          &       & 0.390 & 0.270 & 0.200 & 0.390 & 0.350 & 0.270 & 0.260 & 0.100 & 0.100 & 0.140 & 0.100 & 0.100 & 0.260 & 0.170 & 0.170 \\
\cmidrule{2-17}          & \multirow{2}[2]{*}{\textbf{Point Cov}} & 0.940 & 0.960 & 0.980 & 0.950 & 0.970 & 0.980 & 0.980 & 0.990 & 0.990 & 1.000 & 1.000 & 1.000 & 0.950 & 0.990 & 0.980 \\
          &       & 0.160 & 0.150 & 0.090 & 0.160 & 0.120 & 0.100 & 0.100 & 0.070 & 0.070 & 0.040 & 0.040 & 0.050 & 0.200 & 0.050 & 0.120 \\
\cmidrule{2-17}          & \multirow{3}[2]{*}{\textbf{Amplitude}} & 16.240 & 28.990 & 41.420 & 6.360 & 7.400 & 8.250 & 55.700 & 69.560 & 69.560 & 74.730 & 84.450 & 84.450 & 57.360 & 69.370 & 69.370 \\
          &       & 10.040 & 19.030 & 27.620 & 6.350 & 7.410 & 8.200 & 37.810 & 44.280 & 44.280 & 46.120 & 51.040 & 51.040 & 30.160 & 48.280 & 48.280 \\
          &       & 15.960 & 31.960 & 41.060 & 0.350 & 0.420 & 0.490 & 60.770 & 75.270 & 75.270 & 87.750 & 99.060 & 99.060 & 73.220 & 73.900 & 73.900 \\
\bottomrule[0.3ex]
\end{tabular}%
}
\label{tab.Sim3}%
\end{table}%

\newpage
% Table generated by Excel2LaTeX from sheet 'Gaussian-Wiener'
\begin{table}[htbp]
    \centering
  \caption{Monte Carlo experiment. Setting: asymmetric kernel and Wiener noise. Metrics: Mean and standard deviation for simultaneous coverage and point coverage. Mean, median and standard deviation for amplitude. --\scriptsize Results for different target coverages, band construction methods and sample sizes; 500 MC replications.--}
\renewcommand{\arraystretch}{1} % Tighter table
\scalebox{0.7}{   
    \begin{tabular}{ccccccccccccccccc}
    \toprule
    \multicolumn{1}{c}{\multirow{2}[4]{*}{\textbf{Sample Size}}} & \multirow{2}[4]{*}{\textbf{Metrics}} & \multicolumn{3}{p{15em}}{\textbf{Entropy}} & \multicolumn{3}{p{15em}}{\textbf{GPs}} & \multicolumn{3}{p{15em}}{\textbf{MBD}} & \multicolumn{3}{p{15em}}{\textbf{RPD}} & \multicolumn{3}{p{15em}}{\textbf{L2}} \\
\cmidrule{3-17}          &       & \textbf{80\%} & \textbf{90\%} & \textbf{95\%} & \textbf{80\%} & \textbf{90\%} & \textbf{95\%} & \textbf{80\%} & \textbf{90\%} & \textbf{95\%} & \textbf{80\%} & \textbf{90\%} & \textbf{95\%} & \textbf{80\%} & \textbf{90\%} & \textbf{95\%} \\
    \midrule
    \multirow{7}[6]{*}{\textbf{N = 50}} & \multirow{2}[2]{*}{\textbf{Sim. Cov.}} & 0.740 & 0.780 & 0.850 & 0.540 & 0.680 & 0.750 & 0.760 & 0.860 & 0.920 & 0.860 & 0.890 & 0.920 & 0.670 & 0.840 & 0.940 \\
          &       & 0.440 & 0.420 & 0.360 & 0.500 & 0.470 & 0.440 & 0.430 & 0.350 & 0.270 & 0.350 & 0.310 & 0.270 & 0.470 & 0.370 & 0.240 \\
\cmidrule{2-17}          & \multirow{2}[2]{*}{\textbf{Point Cov}} & 0.910 & 0.940 & 0.970 & 0.790 & 0.870 & 0.920 & 0.940 & 0.950 & 0.980 & 0.960 & 0.980 & 0.980 & 0.910 & 0.960 & 0.990 \\
          &       & 0.190 & 0.150 & 0.090 & 0.310 & 0.260 & 0.220 & 0.160 & 0.160 & 0.100 & 0.130 & 0.100 & 0.090 & 0.190 & 0.130 & 0.070 \\
\cmidrule{2-17}          & \multirow{3}[2]{*}{\textbf{Amplitude}} & 3.350 & 4.380 & 8.090 & 3.930 & 4.680 & 5.300 & 5.290 & 6.420 & 6.420 & 7.910 & 7.760 & 7.760 & 3.690 & 4.780 & 4.780 \\
          &       & 3.150 & 3.910 & 4.750 & 3.780 & 4.520 & 5.180 & 4.120 & 4.550 & 4.550 & 5.020 & 5.190 & 5.190 & 3.000 & 4.420 & 4.420 \\
          &       & 1.790 & 2.980 & 13.730 & 0.830 & 1.060 & 1.230 & 6.520 & 7.860 & 7.860 & 10.880 & 8.970 & 8.970 & 2.170 & 2.430 & 2.430 \\
    \midrule
    \multirow{7}[6]{*}{\textbf{N = 100}} & \multirow{2}[2]{*}{\textbf{Sim. Cov.}} & 0.820 & 0.860 & 0.910 & 0.620 & 0.760 & 0.810 & 0.840 & 0.860 & 0.920 & 0.850 & 0.880 & 0.910 & 0.780 & 0.860 & 0.900 \\
          &       & 0.390 & 0.350 & 0.290 & 0.490 & 0.430 & 0.390 & 0.370 & 0.350 & 0.270 & 0.360 & 0.330 & 0.290 & 0.420 & 0.350 & 0.300 \\
\cmidrule{2-17}          & \multirow{2}[2]{*}{\textbf{Point Cov}} & 0.930 & 0.960 & 0.970 & 0.810 & 0.870 & 0.900 & 0.930 & 0.960 & 0.980 & 0.960 & 0.960 & 0.980 & 0.920 & 0.970 & 0.970 \\
          &       & 0.180 & 0.130 & 0.120 & 0.330 & 0.290 & 0.260 & 0.200 & 0.140 & 0.080 & 0.170 & 0.150 & 0.100 & 0.210 & 0.120 & 0.140 \\
\cmidrule{2-17}          & \multirow{3}[2]{*}{\textbf{Amplitude}} & 3.690 & 4.700 & 6.250 & 4.390 & 5.250 & 5.990 & 13.680 & 16.780 & 16.780 & 21.720 & 22.400 & 22.400 & 6.910 & 15.360 & 15.360 \\
          &       & 3.440 & 4.030 & 4.320 & 4.320 & 5.080 & 5.750 & 6.420 & 9.370 & 9.370 & 9.900 & 14.300 & 14.300 & 4.110 & 6.670 & 6.670 \\
          &       & 1.950 & 2.630 & 6.960 & 0.750 & 0.970 & 1.180 & 23.170 & 24.930 & 24.930 & 37.900 & 32.080 & 32.080 & 10.460 & 28.820 & 28.820 \\
    \midrule
    \multirow{7}[6]{*}{\textbf{N = 250}} & \multirow{2}[2]{*}{\textbf{Sim. Cov.}} & 0.820 & 0.870 & 0.940 & 0.710 & 0.800 & 0.900 & 0.920 & 0.960 & 0.970 & 0.970 & 0.980 & 0.980 & 0.870 & 0.970 & 0.990 \\
          &       & 0.390 & 0.340 & 0.240 & 0.460 & 0.400 & 0.300 & 0.270 & 0.200 & 0.170 & 0.170 & 0.140 & 0.140 & 0.340 & 0.170 & 0.100 \\
\cmidrule{2-17}          & \multirow{2}[2]{*}{\textbf{Point Cov}} & 0.940 & 0.950 & 0.970 & 0.850 & 0.910 & 0.950 & 0.960 & 1.000 & 1.000 & 0.980 & 1.000 & 1.000 & 0.950 & 0.990 & 1.000 \\
          &       & 0.150 & 0.170 & 0.120 & 0.310 & 0.240 & 0.200 & 0.140 & 0.020 & 0.020 & 0.130 & 0.010 & 0.010 & 0.190 & 0.080 & 0.000 \\
\cmidrule{2-17}          & \multirow{3}[2]{*}{\textbf{Amplitude}} & 4.160 & 6.680 & 11.240 & 4.880 & 5.870 & 6.720 & 26.440 & 34.950 & 34.950 & 36.030 & 42.520 & 42.520 & 22.440 & 35.500 & 35.500 \\
          &       & 3.790 & 5.360 & 8.220 & 4.810 & 5.720 & 6.700 & 11.840 & 18.080 & 18.080 & 17.310 & 22.140 & 22.140 & 7.960 & 15.490 & 15.490 \\
          &       & 2.090 & 5.040 & 9.260 & 0.770 & 0.950 & 1.140 & 38.300 & 49.200 & 49.200 & 48.780 & 55.490 & 55.490 & 41.900 & 55.600 & 55.600 \\
    \midrule
    \multirow{7}[6]{*}{\textbf{N = 500}} & \multirow{2}[2]{*}{\textbf{Sim. Cov.}} & 0.860 & 0.950 & 0.970 & 0.790 & 0.900 & 0.960 & 0.880 & 0.960 & 1.000 & 0.980 & 0.970 & 1.000 & 0.880 & 0.960 & 1.000 \\
          &       & 0.350 & 0.220 & 0.170 & 0.410 & 0.300 & 0.200 & 0.330 & 0.200 & 0.000 & 0.140 & 0.170 & 0.000 & 0.330 & 0.200 & 0.000 \\
\cmidrule{2-17}          & \multirow{2}[2]{*}{\textbf{Point Cov}} & 0.960 & 0.990 & 0.990 & 0.910 & 0.960 & 0.980 & 0.960 & 0.990 & 1.000 & 0.990 & 0.990 & 1.000 & 0.950 & 1.000 & 1.000 \\
          &       & 0.130 & 0.050 & 0.070 & 0.220 & 0.150 & 0.100 & 0.120 & 0.080 & 0.000 & 0.060 & 0.060 & 0.000 & 0.170 & 0.020 & 0.000 \\
\cmidrule{2-17}          & \multirow{3}[2]{*}{\textbf{Amplitude}} & 4.990 & 8.530 & 11.780 & 5.010 & 6.050 & 6.860 & 29.670 & 35.000 & 35.000 & 37.960 & 41.320 & 41.320 & 26.840 & 40.620 & 40.620 \\
          &       & 4.070 & 6.610 & 8.020 & 4.940 & 6.020 & 6.860 & 17.510 & 24.040 & 24.040 & 24.080 & 26.830 & 26.830 & 13.750 & 26.690 & 26.690 \\
          &       & 3.430 & 6.570 & 14.880 & 0.500 & 0.630 & 0.750 & 30.160 & 34.510 & 34.510 & 37.960 & 39.580 & 39.580 & 32.730 & 39.730 & 39.730 \\
    \midrule
    \multirow{7}[6]{*}{\textbf{N = 1000}} & \multirow{2}[2]{*}{\textbf{Sim. Cov.}} & 0.840 & 0.910 & 0.970 & 0.820 & 0.890 & 0.930 & 0.920 & 0.950 & 1.000 & 0.950 & 0.990 & 0.990 & 0.920 & 0.990 & 0.990 \\
          &       & 0.370 & 0.290 & 0.170 & 0.390 & 0.310 & 0.260 & 0.270 & 0.220 & 0.000 & 0.220 & 0.100 & 0.100 & 0.270 & 0.100 & 0.100 \\
\cmidrule{2-17}          & \multirow{2}[2]{*}{\textbf{Point Cov}} & 0.950 & 0.980 & 0.990 & 0.890 & 0.950 & 0.970 & 0.980 & 0.990 & 1.000 & 0.990 & 1.000 & 1.000 & 0.970 & 1.000 & 1.000 \\
          &       & 0.120 & 0.090 & 0.030 & 0.280 & 0.180 & 0.140 & 0.090 & 0.080 & 0.000 & 0.060 & 0.040 & 0.020 & 0.140 & 0.000 & 0.050 \\
\cmidrule{2-17}          & \multirow{3}[2]{*}{\textbf{Amplitude}} & 5.320 & 8.950 & 15.810 & 5.040 & 6.090 & 6.990 & 34.270 & 43.080 & 43.080 & 43.930 & 48.750 & 48.750 & 37.660 & 49.310 & 49.310 \\
          &       & 4.730 & 6.850 & 11.210 & 5.020 & 6.090 & 6.920 & 24.030 & 30.150 & 30.150 & 31.040 & 33.820 & 33.820 & 18.910 & 34.740 & 34.740 \\
          &       & 3.190 & 7.770 & 13.710 & 0.420 & 0.560 & 0.680 & 34.970 & 41.830 & 41.830 & 43.830 & 45.750 & 45.750 & 46.930 & 46.400 & 46.400 \\
\bottomrule[0.3ex]
\end{tabular}%
}
\label{tab.Sim4}%
\end{table}%

\newpage
% Table generated by Excel2LaTeX from sheet 'Gaussian-Wiener'
\begin{table}[htbp]
    \centering
  \caption{Monte Carlo experiment. Setting: asymmetric kernel and Laplacian noise. Metrics: Mean and standard deviation for simultaneous coverage and point coverage. Mean, median and standard deviation for amplitude. --\scriptsize Results for different target coverages, band construction methods and sample sizes; 500 MC replications.--}
\renewcommand{\arraystretch}{1} % Tighter table
\scalebox{0.7}{   
    \begin{tabular}{ccccccccccccccccc}
    \toprule
    \multicolumn{1}{c}{\multirow{2}[4]{*}{\textbf{Sample Size}}} & \multirow{2}[4]{*}{\textbf{Metrics}} & \multicolumn{3}{p{15em}}{\textbf{Entropy}} & \multicolumn{3}{p{15em}}{\textbf{GPs}} & \multicolumn{3}{p{15em}}{\textbf{MBD}} & \multicolumn{3}{p{15em}}{\textbf{RPD}} & \multicolumn{3}{p{15em}}{\textbf{L2}} \\
\cmidrule{3-17}          &       & \textbf{80\%} & \textbf{90\%} & \textbf{95\%} & \textbf{80\%} & \textbf{90\%} & \textbf{95\%} & \textbf{80\%} & \textbf{90\%} & \textbf{95\%} & \textbf{80\%} & \textbf{90\%} & \textbf{95\%} & \textbf{80\%} & \textbf{90\%} & \textbf{95\%} \\
    \midrule
    \multirow{7}[6]{*}{\textbf{N = 50}} & \multirow{2}[2]{*}{\textbf{Sim. Cov.}} & 0.760 & 0.810 & 0.910 & 0.550 & 0.690 & 0.860 & 0.800 & 0.850 & 0.960 & 0.850 & 0.930 & 0.950 & 0.760 & 0.820 & 0.930 \\
          &       & 0.430 & 0.390 & 0.290 & 0.500 & 0.460 & 0.350 & 0.400 & 0.360 & 0.200 & 0.360 & 0.260 & 0.220 & 0.430 & 0.390 & 0.260 \\
\cmidrule{2-17}          & \multirow{2}[2]{*}{\textbf{Point Cov}} & 0.920 & 0.960 & 0.970 & 0.790 & 0.880 & 0.920 & 0.940 & 0.950 & 0.980 & 0.960 & 0.980 & 0.980 & 0.920 & 0.950 & 0.990 \\
          &       & 0.180 & 0.120 & 0.130 & 0.330 & 0.290 & 0.240 & 0.190 & 0.170 & 0.140 & 0.130 & 0.110 & 0.100 & 0.190 & 0.170 & 0.050 \\
\cmidrule{2-17}          & \multirow{3}[2]{*}{\textbf{Amplitude}} & 3.470 & 4.180 & 7.050 & 3.840 & 4.540 & 5.150 & 7.520 & 10.300 & 10.300 & 8.520 & 13.130 & 13.130 & 3.540 & 5.170 & 5.170 \\
          &       & 3.050 & 3.640 & 4.580 & 3.830 & 4.470 & 5.040 & 3.920 & 4.680 & 4.680 & 4.440 & 5.360 & 5.360 & 3.230 & 4.670 & 4.670 \\
          &       & 1.750 & 2.280 & 9.030 & 0.710 & 0.900 & 1.060 & 15.910 & 21.250 & 21.250 & 20.590 & 26.270 & 26.270 & 1.760 & 5.350 & 5.350 \\
    \midrule
    \multirow{7}[6]{*}{\textbf{N = 100}} & \multirow{2}[2]{*}{\textbf{Sim. Cov.}} & 0.800 & 0.910 & 0.920 & 0.650 & 0.740 & 0.830 & 0.790 & 0.910 & 0.910 & 0.830 & 0.950 & 0.960 & 0.810 & 0.890 & 0.940 \\
          &       & 0.400 & 0.290 & 0.270 & 0.480 & 0.440 & 0.380 & 0.410 & 0.290 & 0.290 & 0.380 & 0.220 & 0.200 & 0.390 & 0.310 & 0.240 \\
\cmidrule{2-17}          & \multirow{2}[2]{*}{\textbf{Point Cov}} & 0.940 & 0.960 & 0.970 & 0.790 & 0.880 & 0.920 & 0.910 & 0.950 & 0.960 & 0.950 & 0.970 & 0.970 & 0.920 & 0.950 & 0.970 \\
          &       & 0.160 & 0.160 & 0.140 & 0.340 & 0.260 & 0.230 & 0.230 & 0.200 & 0.170 & 0.160 & 0.140 & 0.150 & 0.220 & 0.200 & 0.160 \\
\cmidrule{2-17}          & \multirow{3}[2]{*}{\textbf{Amplitude}} & 3.800 & 5.830 & 7.620 & 4.470 & 5.360 & 6.090 & 9.670 & 21.470 & 21.470 & 17.030 & 26.480 & 26.480 & 7.830 & 14.980 & 14.980 \\
          &       & 3.190 & 4.510 & 5.560 & 4.340 & 5.130 & 5.750 & 4.270 & 9.260 & 9.260 & 5.180 & 10.610 & 10.610 & 3.680 & 6.180 & 6.180 \\
          &       & 2.570 & 4.560 & 6.720 & 0.910 & 1.140 & 1.350 & 13.020 & 34.060 & 34.060 & 29.360 & 36.440 & 36.440 & 16.270 & 19.690 & 19.690 \\
    \midrule
    \multirow{7}[6]{*}{\textbf{N = 250}} & \multirow{2}[2]{*}{\textbf{Sim. Cov.}} & 0.810 & 0.920 & 0.960 & 0.680 & 0.850 & 0.910 & 0.880 & 0.950 & 0.980 & 0.950 & 0.960 & 0.980 & 0.880 & 0.930 & 1.000 \\
          &       & 0.390 & 0.270 & 0.200 & 0.470 & 0.360 & 0.290 & 0.330 & 0.220 & 0.140 & 0.220 & 0.200 & 0.140 & 0.330 & 0.260 & 0.000 \\
\cmidrule{2-17}          & \multirow{2}[2]{*}{\textbf{Point Cov}} & 0.910 & 0.960 & 0.990 & 0.880 & 0.940 & 0.960 & 0.950 & 0.970 & 0.990 & 0.970 & 0.980 & 0.990 & 0.960 & 0.960 & 1.000 \\
          &       & 0.220 & 0.150 & 0.080 & 0.250 & 0.210 & 0.180 & 0.200 & 0.140 & 0.070 & 0.140 & 0.120 & 0.050 & 0.160 & 0.190 & 0.000 \\
\cmidrule{2-17}          & \multirow{3}[2]{*}{\textbf{Amplitude}} & 4.510 & 7.660 & 16.620 & 4.750 & 5.720 & 6.560 & 19.200 & 23.590 & 23.590 & 28.520 & 29.070 & 29.070 & 12.830 & 24.600 & 24.600 \\
          &       & 3.410 & 5.530 & 10.180 & 4.600 & 5.550 & 6.400 & 11.010 & 13.530 & 13.530 & 15.080 & 16.490 & 16.490 & 7.640 & 14.710 & 14.710 \\
          &       & 3.000 & 6.940 & 21.260 & 0.650 & 0.820 & 1.040 & 27.020 & 30.770 & 30.770 & 40.240 & 37.460 & 37.460 & 14.590 & 34.750 & 34.750 \\
    \midrule
    \multirow{7}[6]{*}{\textbf{N = 500}} & \multirow{2}[2]{*}{\textbf{Sim. Cov.}} & 0.860 & 0.930 & 0.980 & 0.770 & 0.860 & 0.920 & 0.860 & 0.900 & 0.980 & 0.940 & 0.960 & 0.980 & 0.780 & 0.930 & 0.980 \\
          &       & 0.350 & 0.260 & 0.140 & 0.420 & 0.350 & 0.270 & 0.350 & 0.300 & 0.140 & 0.240 & 0.200 & 0.140 & 0.420 & 0.260 & 0.140 \\
\cmidrule{2-17}          & \multirow{2}[2]{*}{\textbf{Point Cov}} & 0.960 & 0.970 & 0.990 & 0.890 & 0.950 & 0.980 & 0.920 & 0.950 & 0.980 & 0.950 & 0.980 & 0.990 & 0.900 & 0.960 & 0.990 \\
          &       & 0.140 & 0.120 & 0.090 & 0.250 & 0.160 & 0.130 & 0.250 & 0.180 & 0.120 & 0.200 & 0.120 & 0.110 & 0.240 & 0.150 & 0.080 \\
\cmidrule{2-17}          & \multirow{3}[2]{*}{\textbf{Amplitude}} & 5.320 & 9.070 & 14.460 & 4.940 & 5.920 & 6.790 & 27.700 & 40.010 & 40.010 & 37.430 & 48.240 & 48.240 & 32.020 & 44.500 & 44.500 \\
          &       & 4.250 & 6.800 & 8.850 & 4.900 & 5.860 & 6.720 & 19.150 & 25.610 & 25.610 & 23.450 & 35.050 & 35.050 & 13.900 & 28.540 & 28.540 \\
          &       & 4.680 & 7.990 & 18.040 & 0.520 & 0.610 & 0.750 & 30.010 & 44.220 & 44.220 & 39.330 & 48.390 & 48.390 & 58.140 & 51.570 & 51.570 \\
    \midrule
    \multirow{7}[6]{*}{\textbf{N = 1000}} & \multirow{2}[2]{*}{\textbf{Sim. Cov.}} & 0.880 & 0.920 & 0.970 & 0.840 & 0.940 & 0.990 & 0.950 & 0.970 & 0.990 & 0.980 & 0.980 & 0.990 & 0.910 & 0.990 & 0.990 \\
          &       & 0.330 & 0.270 & 0.170 & 0.370 & 0.240 & 0.100 & 0.220 & 0.170 & 0.100 & 0.140 & 0.140 & 0.100 & 0.290 & 0.100 & 0.100 \\
\cmidrule{2-17}          & \multirow{2}[2]{*}{\textbf{Point Cov}} & 0.970 & 0.970 & 0.990 & 0.940 & 0.980 & 0.990 & 0.980 & 0.980 & 1.000 & 0.990 & 0.990 & 1.000 & 0.940 & 0.990 & 1.000 \\
          &       & 0.100 & 0.110 & 0.090 & 0.180 & 0.100 & 0.100 & 0.130 & 0.110 & 0.050 & 0.100 & 0.090 & 0.010 & 0.220 & 0.100 & 0.030 \\
\cmidrule{2-17}          & \multirow{3}[2]{*}{\textbf{Amplitude}} & 5.750 & 9.080 & 17.580 & 5.020 & 6.060 & 6.930 & 35.390 & 42.470 & 42.470 & 48.070 & 52.560 & 52.560 & 33.770 & 52.480 & 52.480 \\
          &       & 4.880 & 6.680 & 10.970 & 5.020 & 5.970 & 6.820 & 21.370 & 31.780 & 31.780 & 30.930 & 36.350 & 36.350 & 18.640 & 34.080 & 34.080 \\
          &       & 3.920 & 8.770 & 19.080 & 0.410 & 0.510 & 0.650 & 39.480 & 41.790 & 41.790 & 52.910 & 52.120 & 52.120 & 43.760 & 58.330 & 58.330 \\
\bottomrule[0.3ex]
\end{tabular}%
}
\label{tab.Sim5}%
\end{table}%

\newpage
% Table generated by Excel2LaTeX from sheet 'Gaussian-Wiener'
\begin{table}[htbp]
    \centering
  \caption{Monte Carlo experiment. Setting: asymmetric kernel and exponential noise. Metrics: Mean and standard deviation for simultaneous coverage and point coverage. Mean, median and standard deviation for amplitude. --\scriptsize Results for different target coverages, band construction methods and sample sizes; 500 MC replications.--}
\renewcommand{\arraystretch}{1} % Tighter table
\scalebox{0.7}{   
    \begin{tabular}{ccccccccccccccccc}
    \toprule
    \multicolumn{1}{c}{\multirow{2}[4]{*}{\textbf{Sample Size}}} & \multirow{2}[4]{*}{\textbf{Metrics}} & \multicolumn{3}{p{15em}}{\textbf{Entropy}} & \multicolumn{3}{p{15em}}{\textbf{GPs}} & \multicolumn{3}{p{15em}}{\textbf{MBD}} & \multicolumn{3}{p{15em}}{\textbf{RPD}} & \multicolumn{3}{p{15em}}{\textbf{L2}} \\
\cmidrule{3-17}          &       & \textbf{80\%} & \textbf{90\%} & \textbf{95\%} & \textbf{80\%} & \textbf{90\%} & \textbf{95\%} & \textbf{80\%} & \textbf{90\%} & \textbf{95\%} & \textbf{80\%} & \textbf{90\%} & \textbf{95\%} & \textbf{80\%} & \textbf{90\%} & \textbf{95\%} \\
    \midrule
    \multirow{7}[6]{*}{\textbf{N = 50}} & \multirow{2}[2]{*}{\textbf{Sim. Cov.}} & 0.710 & 0.830 & 0.960 & 0.610 & 0.760 & 0.850 & 0.750 & 0.850 & 0.930 & 0.890 & 0.930 & 0.950 & 0.750 & 0.810 & 0.950 \\
          &       & 0.460 & 0.380 & 0.200 & 0.490 & 0.430 & 0.360 & 0.440 & 0.360 & 0.260 & 0.310 & 0.260 & 0.220 & 0.440 & 0.390 & 0.220 \\
\cmidrule{2-17}          & \multirow{2}[2]{*}{\textbf{Point Cov}} & 0.920 & 0.960 & 0.990 & 0.820 & 0.900 & 0.930 & 0.920 & 0.960 & 0.970 & 0.970 & 0.980 & 0.980 & 0.920 & 0.960 & 0.990 \\
          &       & 0.180 & 0.130 & 0.070 & 0.310 & 0.240 & 0.200 & 0.170 & 0.130 & 0.120 & 0.100 & 0.100 & 0.100 & 0.200 & 0.110 & 0.060 \\
\cmidrule{2-17}          & \multirow{3}[2]{*}{\textbf{Amplitude}} & 9.960 & 14.980 & 29.210 & 7.820 & 9.300 & 10.550 & 44.820 & 38.700 & 38.700 & 66.660 & 57.100 & 57.100 & 13.100 & 20.290 & 20.290 \\
          &       & 6.100 & 7.580 & 13.670 & 7.520 & 8.790 & 10.000 & 11.110 & 15.910 & 15.910 & 17.640 & 23.860 & 23.860 & 6.580 & 7.960 & 7.960 \\
          &       & 17.040 & 23.410 & 49.160 & 1.630 & 2.040 & 2.400 & 168.430 & 52.500 & 52.500 & 213.980 & 84.730 & 84.730 & 24.800 & 33.500 & 33.500 \\
    \midrule
    \multirow{7}[6]{*}{\textbf{N = 100}} & \multirow{2}[2]{*}{\textbf{Sim. Cov.}} & 0.900 & 0.960 & 0.960 & 0.620 & 0.780 & 0.850 & 0.880 & 0.930 & 0.950 & 0.940 & 0.960 & 0.960 & 0.800 & 0.900 & 0.930 \\
          &       & 0.300 & 0.200 & 0.200 & 0.490 & 0.420 & 0.360 & 0.330 & 0.260 & 0.220 & 0.240 & 0.200 & 0.200 & 0.400 & 0.300 & 0.260 \\
\cmidrule{2-17}          & \multirow{2}[2]{*}{\textbf{Point Cov}} & 0.980 & 0.990 & 0.990 & 0.810 & 0.880 & 0.910 & 0.950 & 0.980 & 0.980 & 0.980 & 0.980 & 0.980 & 0.920 & 0.960 & 0.980 \\
          &       & 0.090 & 0.100 & 0.050 & 0.340 & 0.290 & 0.250 & 0.160 & 0.110 & 0.110 & 0.100 & 0.100 & 0.080 & 0.220 & 0.170 & 0.090 \\
\cmidrule{2-17}          & \multirow{3}[2]{*}{\textbf{Amplitude}} & 9.560 & 17.790 & 27.560 & 8.720 & 10.440 & 11.900 & 46.880 & 57.450 & 57.450 & 77.840 & 80.400 & 80.400 & 23.500 & 54.050 & 54.050 \\
          &       & 6.020 & 10.310 & 13.460 & 8.590 & 10.300 & 11.660 & 20.750 & 27.100 & 27.100 & 29.730 & 37.150 & 37.150 & 9.810 & 22.600 & 22.600 \\
          &       & 11.620 & 17.180 & 40.640 & 1.590 & 2.020 & 2.360 & 57.280 & 66.890 & 66.890 & 130.750 & 117.820 & 117.820 & 37.750 & 88.130 & 88.130 \\
    \midrule
    \multirow{7}[6]{*}{\textbf{N = 250}} & \multirow{2}[2]{*}{\textbf{Sim. Cov.}} & 0.910 & 0.970 & 0.980 & 0.700 & 0.850 & 0.890 & 0.860 & 0.960 & 0.960 & 0.990 & 0.960 & 0.990 & 0.910 & 0.940 & 0.990 \\
          &       & 0.290 & 0.170 & 0.140 & 0.460 & 0.360 & 0.310 & 0.350 & 0.200 & 0.200 & 0.100 & 0.200 & 0.100 & 0.290 & 0.240 & 0.100 \\
\cmidrule{2-17}          & \multirow{2}[2]{*}{\textbf{Point Cov}} & 0.960 & 0.990 & 0.990 & 0.870 & 0.930 & 0.960 & 0.950 & 0.990 & 0.990 & 1.000 & 1.000 & 1.000 & 0.970 & 0.970 & 1.000 \\
          &       & 0.160 & 0.060 & 0.060 & 0.280 & 0.200 & 0.140 & 0.160 & 0.040 & 0.040 & 0.020 & 0.030 & 0.030 & 0.140 & 0.160 & 0.010 \\
\cmidrule{2-17}          & \multirow{3}[2]{*}{\textbf{Amplitude}} & 11.100 & 26.600 & 44.290 & 9.670 & 11.610 & 13.240 & 39.120 & 57.800 & 57.800 & 67.720 & 72.230 & 72.230 & 40.610 & 57.720 & 57.720 \\
          &       & 8.000 & 14.800 & 29.670 & 9.440 & 11.400 & 12.880 & 27.260 & 48.070 & 48.070 & 52.950 & 58.890 & 58.890 & 20.100 & 42.460 & 42.460 \\
          &       & 9.290 & 36.880 & 44.610 & 1.400 & 1.670 & 2.010 & 37.310 & 52.460 & 52.460 & 69.710 & 64.150 & 64.150 & 46.900 & 50.630 & 50.630 \\
    \midrule
    \multirow{7}[6]{*}{\textbf{N = 500}} & \multirow{2}[2]{*}{\textbf{Sim. Cov.}} & 0.900 & 0.970 & 0.950 & 0.770 & 0.870 & 0.920 & 0.890 & 0.960 & 0.980 & 0.970 & 0.970 & 0.990 & 0.880 & 0.970 & 0.970 \\
          &       & 0.300 & 0.170 & 0.220 & 0.420 & 0.340 & 0.270 & 0.310 & 0.200 & 0.140 & 0.170 & 0.170 & 0.100 & 0.330 & 0.170 & 0.170 \\
\cmidrule{2-17}          & \multirow{2}[2]{*}{\textbf{Point Cov}} & 0.950 & 0.980 & 0.980 & 0.900 & 0.950 & 0.960 & 0.960 & 0.980 & 0.990 & 0.990 & 0.990 & 0.990 & 0.940 & 0.990 & 0.980 \\
          &       & 0.160 & 0.090 & 0.090 & 0.250 & 0.190 & 0.160 & 0.140 & 0.110 & 0.080 & 0.090 & 0.090 & 0.070 & 0.210 & 0.080 & 0.100 \\
\cmidrule{2-17}          & \multirow{3}[2]{*}{\textbf{Amplitude}} & 16.390 & 26.950 & 44.530 & 9.990 & 11.990 & 13.660 & 44.050 & 63.140 & 63.140 & 67.260 & 82.750 & 82.750 & 47.110 & 64.710 & 64.710 \\
          &       & 9.610 & 14.600 & 20.380 & 9.890 & 11.870 & 13.460 & 22.900 & 37.440 & 37.440 & 36.610 & 51.990 & 51.990 & 20.840 & 36.340 & 36.340 \\
          &       & 28.210 & 37.270 & 53.750 & 1.200 & 1.520 & 1.740 & 58.770 & 69.620 & 69.620 & 84.710 & 91.080 & 91.080 & 75.520 & 71.890 & 71.890 \\
    \midrule
    \multirow{7}[6]{*}{\textbf{N = 1000}} & \multirow{2}[2]{*}{\textbf{Sim. Cov.}} & 0.890 & 0.920 & 0.930 & 0.750 & 0.860 & 0.930 & 0.910 & 0.860 & 0.830 & 0.950 & 0.930 & 0.930 & 0.880 & 0.910 & 0.900 \\
          &       & 0.310 & 0.270 & 0.260 & 0.440 & 0.350 & 0.260 & 0.290 & 0.350 & 0.380 & 0.220 & 0.260 & 0.260 & 0.330 & 0.290 & 0.300 \\
\cmidrule{2-17}          & \multirow{2}[2]{*}{\textbf{Point Cov}} & 0.960 & 0.970 & 0.970 & 0.870 & 0.940 & 0.960 & 0.950 & 0.920 & 0.900 & 0.970 & 0.970 & 0.970 & 0.930 & 0.930 & 0.930 \\
          &       & 0.160 & 0.130 & 0.130 & 0.290 & 0.200 & 0.160 & 0.200 & 0.240 & 0.260 & 0.160 & 0.150 & 0.150 & 0.230 & 0.220 & 0.230 \\
\cmidrule{2-17}          & \multirow{3}[2]{*}{\textbf{Amplitude}} & 16.050 & 29.860 & 38.180 & 10.110 & 12.260 & 13.990 & 54.180 & 45.760 & 45.760 & 74.160 & 58.080 & 58.080 & 58.490 & 58.260 & 58.260 \\
          &       & 10.920 & 17.310 & 22.320 & 10.090 & 12.110 & 13.700 & 37.280 & 28.110 & 28.110 & 55.720 & 36.560 & 36.560 & 33.520 & 33.500 & 33.500 \\
          &       & 15.800 & 38.600 & 45.180 & 0.930 & 1.240 & 1.370 & 51.040 & 48.080 & 48.080 & 67.680 & 58.080 & 58.080 & 61.030 & 60.660 & 60.660 \\
\bottomrule[0.3ex]
\end{tabular}%
}
\label{tab.Sim6}%
\end{table}%
\end{landscape}

\bibliography{references}

\end{document}